\documentclass[aps,pra,superscriptaddress,notitlepage,twocolumn]{revtex4-2}
\pdfoutput=1 
\usepackage[utf8]{inputenc}
\usepackage{braket}
\usepackage{amsfonts}
\usepackage{amsthm}
\usepackage{amssymb}
\usepackage{graphicx}
\usepackage{bm}
\usepackage{bbm}
\usepackage{xcolor}
\usepackage{comment}
\usepackage{mathtools}
\usepackage{physics}
\usepackage{enumitem}

\newcommand{\ba}{\begin{eqnarray}}
\newcommand{\ea}{\end{eqnarray}}
\newcommand{\ban}{\begin{eqnarray*}}
\newcommand{\ean}{\end{eqnarray*}}
\newcommand{\one}{\mathbbm{1}}
\DeclareMathOperator{\EE}{\mathbb{E}}
\DeclareMathOperator{\sgn}{sgn}
\DeclareMathOperator{\pos}{pos}
\allowdisplaybreaks

\begin{document}

\title{Detecting quantumness in uniform precessions}
\author{Lin Htoo Zaw}
\affiliation{Centre for Quantum Technologies, National University of Singapore, 3 Science Drive 2, Singapore 117543}
\author{Clive Cenxin Aw}
\affiliation{Centre for Quantum Technologies, National University of Singapore, 3 Science Drive 2, Singapore 117543}
\author{Zakarya Lasmar}
\affiliation{Centre for Quantum Technologies, National University of Singapore, 3 Science Drive 2, Singapore 117543}
\author{Valerio Scarani}
\affiliation{Centre for Quantum Technologies, National University of Singapore, 3 Science Drive 2, Singapore 117543}
\affiliation{Department of Physics, National University of Singapore, 2 Science Drive 3, Singapore 117542}

\begin{abstract}
Building on work by Tsirelson, we present a family of protocols that detect the nonclassicality of suitable states of a single quantum system, under the sole assumption that the measured dynamical observable undergoes a uniform precession. The case of the harmonic oscillator was anticipated in the work by Tsirelson, which we extend. We then apply the protocols to finite-dimensional spins that undergo uniform precession in real space and find a gap between the classical and the quantum expectations for every $j\geq \frac{3}{2}$ (excluding $j=2$).
\end{abstract}

\maketitle

\section{Introduction}

Uniform precession is one of the most basic forms of motion. If a Hamiltonian $H$ generates uniform precession of some dynamical variables in classical theory, the quantum theory obtained by canonical quantization also predicts uniform precession of the corresponding observables in the Heisenberg representation.

The one-dimensional harmonic oscillator is paradigmatic. Imported into quantum theory by Heisenberg, as a side remark while tackling the nonharmonic oscillator \cite{Heis25}, it is now textbook knowledge, in the even simpler algebraic approach of Dirac \cite{diracbook}. Since Planck's blackbody radiation and the Einstein-Debye specific heat, countless observations have been made of nonclassical features of harmonic systems. Today's level of control over individual quantum harmonic systems extends from optical modes as the most explored platform \cite{HRbook} to mechanical modes as the frontier \cite{entg18,entg21,levito}. All the reported quantum features of harmonic oscillators have to do with the kinematics (preparing nonclassical states, which are actually independent of the dynamics), or originate from the discrete nature of the energy spectrum (e.g.~the presence of a zero-point energy). By contrast, the dynamics of the independent variables $(x,p)$ is a uniform precession in phase space both in classical and in quantum theory. It is then thought that proper quantum features of harmonic oscillators have to be found elsewhere.

Building on a virtually unnoticed work by Tsirelson \cite{tsirelson}, we show that this common belief is unwarranted. For suitable initial states, evidence of nonclassicality can be obtained from the statistics of one measurement, under the sole assumption that the observed dynamical variable undergoes a uniform precession. We extend Tsirelson's observation to systems with finitely many levels (spins). A necessary condition to have a gap between the classical and the quantum prediction is that the precession couples noncommuting variables. The protocols introduced here constitute a different type of certification of nonclassicality.

In Sec.~\ref{sec:classical} we introduce the protocol for general uniformly precessing observables and derive the classical bounds. In Sec.~\ref{sec:ho} we apply the protocol to the same physical system as Tsirelson: precession in the phase space of the one-dimensional harmonic oscillator. In Sec.~\ref{sec:spin} we consider precessions in real space due to spin angular momentum. Sec.~\ref{sec:analytical} sketches some of the analytical results, whose detailed description and derivation are given in the Appendixes.

\section{\label{sec:classical}Classical bounds of uniformly-precessing observables}
Consider two physical quantities $A_1$ and $A_2$ that define a precession, that is, their evolution in time is given by
\begin{equation}
\begin{aligned}
    A_1(t) &= A_1(0)\cos(\omega t) +  A_2(0)\sin(\omega t)\\
    A_2(t) &= A_2(0)\cos(\omega t) -  A_1(0)\sin(\omega t),
\end{aligned}
\end{equation}
for a fixed pulsation $\omega = 2\pi/T$. For classical systems, $A_1(t)$ and $A_2(t)$ are the values of the physical quantities measured at times $t$; for quantum systems, $A_1(t)$ and $A_2(t)$ are their corresponding Hermitian operators given in the Heisenberg representation.

Divide one period $T$ of the precession into $K$ equal times $t_k=(k/K)T$, with $K>0$ an integer and $k=0,1,...,K-1$. In every round of the protocol, one of the times $t_k$ is chosen with uniform probability $1/K$, and the observable $A_1$ is measured at that time. After several rounds, one estimates the average probability of having found $A_1>0$:
\begin{equation}\label{defPN}
\begin{aligned}
{P}_K& \equiv \frac{1}{K}\sum_{k=0}^{K-1}\Big[
\textrm{Prob}(A_1>0 \textrm{ at } t=t_k)\\[-1ex]
&\qquad\qquad\qquad+\tfrac{1}{2}\textrm{Prob}(A_1=0 \textrm{ at } t=t_k)\Big] \\
&= \frac{1}{K}\sum_{k=0}^{K-1} \pos(A_1(t_k)) \,,
\end{aligned}
\end{equation}
where $\pos(x) = \frac{1}{2}(1+\sgn(x))$ is the Heaviside step function. The term $\frac{1}{2}\textrm{Prob}(A_1=0)$ is needed to control the behavior at points $(A_1=0,A_2)$, as we will see shortly.

We will denote by $\mathbf{P}_K$ the maximum of $P_K$ over all states. The results obtained with classical mechanics are denoted by ${P}_K^c$ and $\mathbf{P}^c_K$.

In classical physics, a pure state, i.e., a state of maximal knowledge, is represented by a point $(A_1(t),A_2(t))$ in the $A_1$-$A_2$ plane, which precesses therefore with period $T$. Let us introduce the angle $\theta \equiv \operatorname{arctan2}(A_2,A_1)$ and the open sectors
\begin{equation}
\begin{aligned}
    \Theta_{+k} &= \Big\{(A_1,A_2):\theta \in \left(\tfrac{\pi}{2}, \tfrac{\pi}{2} - \tfrac{\pi}{K} \right) - \tfrac{2\pi k}{K}  \Big\}\\
    \Theta_{-k} &= \Big\{(A_1,A_2):\theta \in  \left(\tfrac{\pi}{2} - \tfrac{\pi}{K}, \tfrac{\pi}{2} - \tfrac{2\pi}{K} \right) - \tfrac{2\pi k}{K} \Big\}\,.
\end{aligned}
\end{equation}
Due to the uniform precession, points initially in the region $\Theta_{\pm k_0}$ will be in the region $\Theta_{\pm \bqty{(k_0 + k) \bmod K}}$ at $t=t_k$.

\begin{figure}[h!]
\includegraphics[width=\columnwidth]{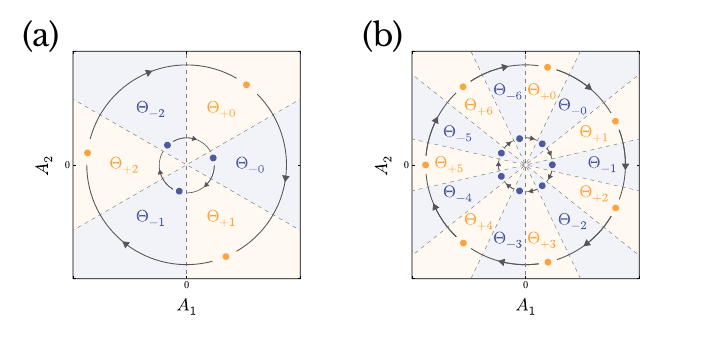}
\caption{\label{fig:classical-case}Classical phase-space picture of the protocol, for (a) $K=3$ and (b) $K=7$, split into regions labeled $\Theta_{\pm k}$, with $k=0,1,\dots,K-1$. The trajectory of several example states are shown. Points in the region $\Theta_{\pm k_0}$ at $t=0$ will be in the region $\Theta_{\pm \bqty{(k_0 + k) \bmod K}}$ at $t=t_k$. For odd $K$, points in the region $\Theta_{+k}$ ($\Theta_{-k}$) have the classical score $P_K^c=\frac{1}{2}(1+\frac{1}{K})$ ($P_K^c=\frac{1}{2}(1-\frac{1}{K})$), which can be easily found by counting how often the point stays on the positive $A_1$ plane. A similar argument for even $K$ gives $P_K^c=\frac{1}{2}$.}
\end{figure}

A simple graphical argument [see Figs.~\ref{fig:classical-case}(a) and \ref{fig:classical-case}(b) for $K=3$ and $K=7$, respectively] tells us that for even $K$, points initially in $\Theta_{\pm k}$ have $A_1$ positive $K/2$ out of $K$ times, giving $P_K^c = \frac{1}{2}$. Meanwhile, for odd $K$, points initially in $\Theta_{\pm k}$ will have $x$ positive $(K \pm 1)/2$ out of $K$ times, giving $P_K^c = \frac{1}{2} \pm \frac{1}{2K}$. As for the points that lie at the boundary between two open sectors, they yield $P_K^c = \frac{1}{2}$; this is the case notably for $(A_1,A_2)=(0,0)$ due to the term $\frac{1}{2}\textrm{Prob}(A_1=0)$ in \eqref{defPN}. Therefore, for a generic classical state,~i.e. for any probability density $\rho(A_1,A_2)$ over phase space, the classical score satisfies
\ba\label{classical}
\frac{1}{2}\left(1- \frac{1}{K}\right)\leq {P}_K^c \leq \frac{1}{2}\left(1+ \frac{1}{K}\right)\equiv \mathbf{P}_K^c&&\;\textrm{for $K$ odd.}
\ea

\begin{figure*}[t!]
\includegraphics[width=\textwidth]{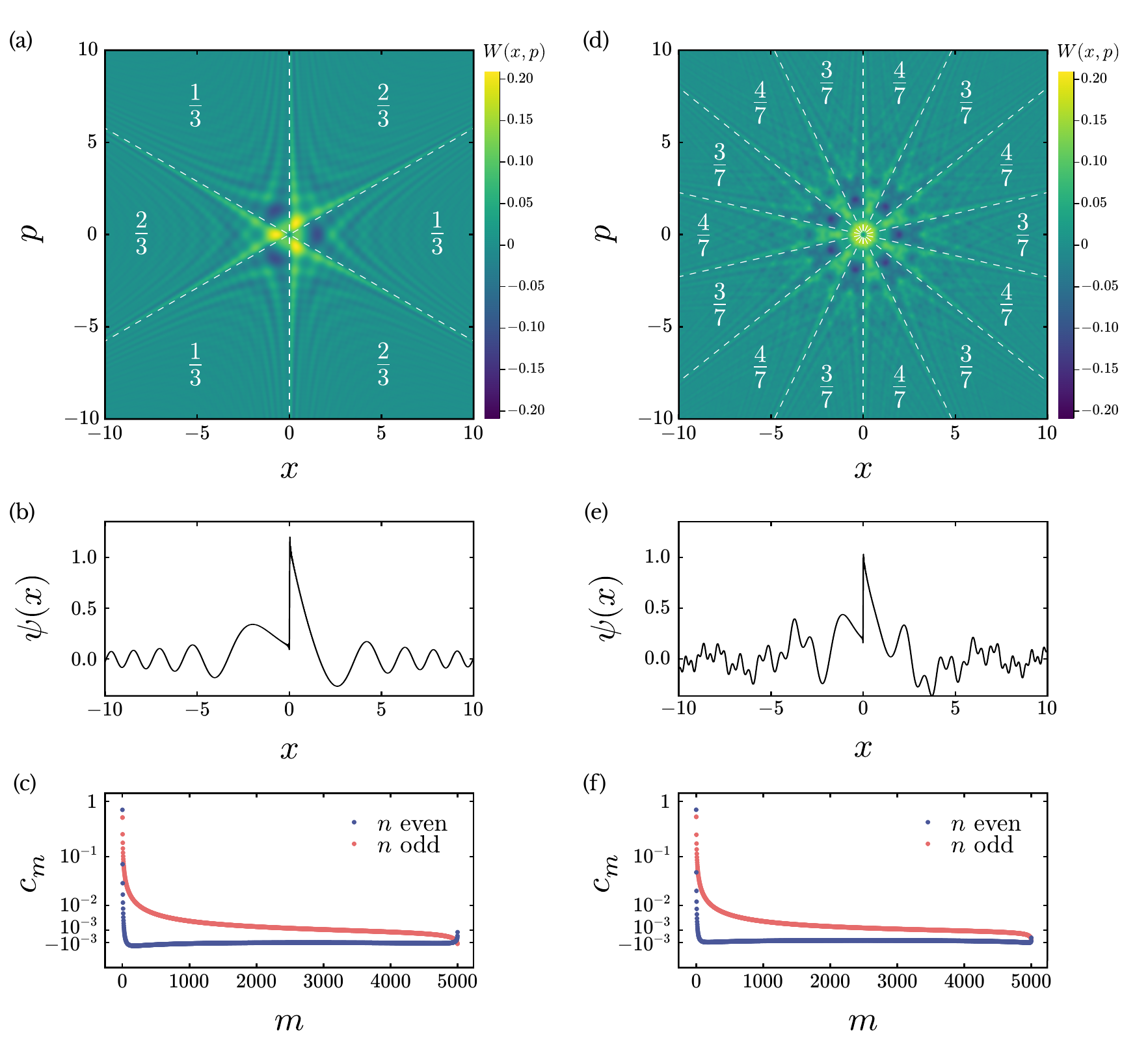}
\caption{\label{fig:wigner-plot}
(a) and (d) Wigner function $W(x,p)$ of the state of the one-dimensional harmonic oscillator that achieves $\mathbf{P}_K^\infty$ (truncated at the energy eigenstate $n=2100$) for (a) $K=3$ and (d) $K=7$. Here $x$ and $p$ are in units of $\sqrt{\hbar/m\omega}$ and $\sqrt{\hbar m\omega}$, respectively. Superimposed are the phase-space sectors and corresponding value of the classical score ${P}_K^c$. Notice how the negativity of $W(x,p)$ is concentrated in the sectors where ${P}_K^c$ is minimal. (b) and (e) Corresponding wave functions $\psi(x)$ (truncated at $n=5000K$). For both the Wigner and wave functions, if $\ket{\psi_n}$ is the state that achieves $\mathbf{P}_K^\infty$ truncated at $n$, then $n$ is chosen so that $\abs{\braket{\psi_n}{\psi_{2n}}}^2 > 0.99$. (c) and (f) The state that achieves $\mathbf{P}_K^\infty$ is of the form $\ket{\psi} = \sum_m  (-1)^{\lceil m/2 \rceil} c_m \ket{mK}$ with $c_m \in \mathbbm{R}$. The coefficients $c_m$, found numerically with the truncation $n \leq 5000K$, are plotted above.}
\end{figure*}

\section{Harmonic oscillator: precession in phase space}\label{sec:ho}
A one-dimensional material point is a system, whose only independent dynamical variables are position $x$ and momentum $p$. This system is called a harmonic oscillator if its dynamics is generated by the Hamiltonian $H=\frac{1}{2m}(p^2+m\omega^2x^2)$. Under this Hamiltonian, $x$ and $p$ evolve in time according to $x(t)=x(0)\cos\omega t + \tilde{p}(0)\sin\omega t$ and $\tilde{p}(t)=\tilde{p}(0)\cos\omega t - x(0)\sin\omega t$, with $\tilde{p}=p/m\omega$. In other words, the evolution describes a precession in phase space.

Therefore, we can identify $(A_1,A_2) = (x,\tilde{p})$ and perform the protocol as introduced in the preceding section. The scores obtained with the quantum harmonic oscillator are denoted by ${P}_K^\infty$ and $\mathbf{P}^\infty_K$ because we are dealing with an infinite-dimensional vector space.

This protocol with $K=3$ was originally considered by Tsirelson \cite{tsirelson}. He
found that $\mathbf{P}_3^\infty\gtrsim 0.709>\mathbf{P}_3^c=\frac{2}{3}$. The origin of such a gap between the quantum and the classical prediction must be traced to the fact that the precession couples incompatible variables. Every probability density satisfies the classical bound: to violate the bound, the joint distribution of $(x,p)$ must be a quasi-probability density with negative values. A very compelling image of Tsirelson's discovery is indeed obtained by plotting the Wigner function $W(x,p)$ of the state that is numerically found to yield $\mathbf{P}_3^\infty$ [Figs.~\ref{fig:wigner-plot}(a)--\ref{fig:wigner-plot}(c)]. Anticipating the use of our techniques, to be described below, we also plot the state for $K=7$, for which $\mathbf{P}_7^\infty\gtrsim 0.61>\mathbf{P}_7^c=\frac{4}{7}$ [Figs.~\ref{fig:wigner-plot}(d)--\ref{fig:wigner-plot}(f)]. The fact that these Wigner functions have the same symmetry as the classical probabilities will be proved rigorously below, where it will also be used to derive an upper bound on $\mathbf{P}_K^\infty$.

\begin{figure*}[ht]
    \includegraphics[width=\textwidth]{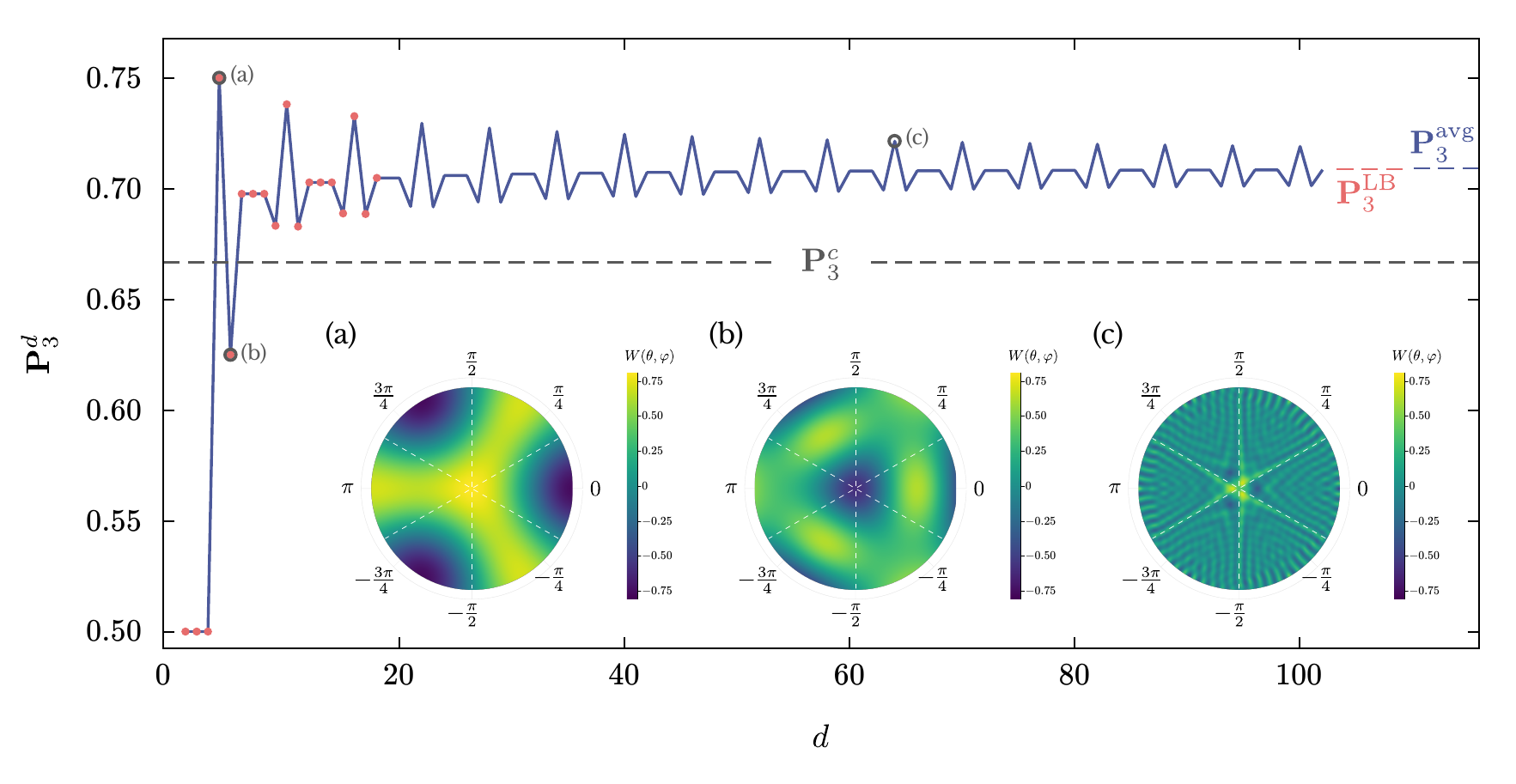}
    \caption{\label{fig:wiggly-3} Values whose analytical expression is known for every $K$ are given in red. The average of $\mathbf{P}_3^d$ over $2000 \leq d \leq 4000$ yields $\mathbf{P}_3^{\text{avg}}=0.7092$; the analytical asymptotic lower bound \eqref{lowerbound} is $\mathbf{P}_3^{\text{LB}}=0.7087$. We find $\mathbf{P}_3^{d} > \mathbf{P}_3^{c}$ for $d=3+1$ and $d \geq 2 \times 3$. Insets: Wigner functions $W(\theta,\varphi)$ defined on spherical phase space \cite{spherical-phase-space} for (a) $d=3+1$, (b) $d=3+2$, and (c) $d=21\times 3 + 1$. The time evolution under the Hamiltonian $H=\omega J_z$ corresponds to a rotation of the Wigner function around the $z$ axis with angular frequency $\omega$. A stereographic projection of its upper hemisphere is shown, where the origin corresponds to $\theta=0$, the boundary corresponds to $\theta=\pi/2$, and the ticks around the boundary label azimuthal angles $\varphi \in [0,2\pi]$. The lower hemisphere ($\theta \in [\pi/2,\pi]$) is an exact reflection of the upper hemisphere ($\theta \in [0,\pi/2]$). Dashed lines mark the same phase-space sectors as in Fig.~\ref{fig:wigner-plot}. Note the similarity between the image in (c) and Fig.~\ref{fig:wigner-plot}(a).
  }
\end{figure*} 

\begin{figure}[b!]
  \includegraphics[width=\columnwidth]{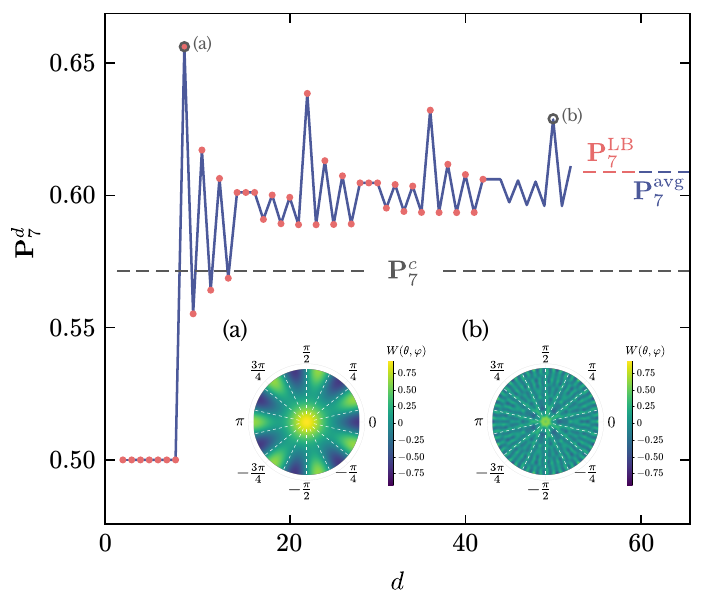}
  \caption{\label{fig:wiggly-7} Same as in Fig.~\ref{fig:wiggly-3} for $K=7$. The average of $\mathbf{P}_7^d$ over $6000 \leq d \leq 8000$ yields $\mathbf{P}_7^{\text{avg}}=0.6088$; the analytical asymptotic lower bound \eqref{lowerbound} is $\mathbf{P}_7^{\text{LB}}=0.6089$. We find $\mathbf{P}_7^{d} > \mathbf{P}_7^{c}$ for $d=7+k$ with $k\in\{1,3,5\}$ and $d \geq 2 \time 7$. The insets show the Wigner function defined on spherical phase space for (a) $d=7+1$ and (b) $d=7\times7+1$. Note also the similarity between the image in (b) and Fig.~\ref{fig:wigner-plot}(d).}
\end{figure}

\section{Spin $j$: precession in real space}\label{sec:spin}
Under the Hamiltonian $H=-\omega J_z$, where $J_z$ is the component of the angular momentum $\vec{J}$ of the system in the $z$ direction, any vector $\vec{V}$ precesses around the $z$ axis at pulsation $\omega$: $V_x(t)=V_x(0)\cos\omega t + V_y(0)\sin\omega t$, and $V_y(t)=V_y(0)\cos\omega t - V_x(0)\sin\omega t$, $V_z(t)=V_z(0)$. Hence, the protocol can be performed by identifying $(A_1,A_2) = (V_x,V_y)$. To have a quantum gap, we need a vector whose quantum description is such that $[V_x,V_y]\neq 0$ (for instance, the position vector $\vec{x}$ would not do). 

For this study, we choose $\vec{J}$ itself, since famously $[J_x,J_y]=i\hbar J_z$. For simplicity of notation, we assume in what follows that $\vec{J}$ is a spin $j$ (the results are of course unchanged for a subspace of fixed $j$ of a more complex angular momentum), with Hilbert space dimension $d=2j+1$. Analogously to the preceding section, the scores obtained with the $d$-dimensional Hilbert space are denoted by ${P}_K^d$ and $\mathbf{P}^d_K$.

We label $\ket{m_x}$ the eigenvector of $J_x$ for the eigenvalue $\hbar m$, with $m\in\{-j,-j+1,...,j-1, j\}$. The detection of a positive component of $J_x$ is then represented by the operator
\ba
\pos(J_x)
&=& \frac{1}{2}\bqty{\mathbbm{1} + \sgn(J_x)},
\ea 
where \ba\sgn(J_x) = \sum_{m=-j}^j \sgn(m)\ketbra{m_x}\,.\ea
Then, given a state, ${P}_K^d$ will be the expectation value of the operator
\ba\label{QK}
Q_K&=&\frac{1}{K}\sum_{k=0}^{K-1} e^{-i\theta_kJ_z}\,\pos(J_x)\, e^{i\theta_kJ_z}\,\equiv \mathbb{E}_K[\pos(J_x)],
\ea with $\theta_k=2\pi k/K$. Thus, the maximum $\mathbf{P}_K^d$ is the largest eigenvalue of $Q_K$ and the optimal state is the corresponding eigenvector.

For $K$ even, $Q_K=\frac{1}{2}\one$ for all $d$. Indeed, we see that $e^{-i\theta_k J_z}\ket{m_x}\equiv \ket{m_{\theta_k}}$, the eigenvector of $\cos\theta_k J_x+\sin\theta_k J_y$ for eigenvalue $m$. However, if $K$ is even, $\theta_{k+K/2}=\theta_k+\pi$ is also in the sum and $e^{-i(\theta_k+\pi) J_z}\ket{m_x}\equiv \ket{-m_{\theta_k}}$. Therefore, for $K$ even, ${P}_K^d=\mathbf{P}_K^d=\mathbf{P}_K^c=\frac{1}{2}$ for every state.

For $K$ odd, finding eigenvalues and eigenvectors of finite matrices can be reliably achieved with standard numerical tools. The results are shown in Fig.~\ref{fig:wiggly-3} for $K=3$ and Fig.~\ref{fig:wiggly-7} for $K=7$; we plotted the same graphs for other odd values of $K$ as well. They all show a systematic behavior of $\mathbf{P}_K^{d}$ as a function of $d$. The first nontrivial value $\mathbf{P}_K^{K+1}$ is also the largest; for larger $d$, the graph exhibits a pattern with periodicity $2K$, while settling around a value that is clearly higher than the classical one. The fact that this value seems to approach $\mathbf{P}_K^{\infty}$ obtained with the harmonic oscillator is not a coincidence, as we are going to discuss below.

\section{Analytical results}
\label{sec:analytical}

The starting point to obtain analytical results is the observation that
\ba\mathbb{E}_K[\ketbra{m_z}{m'_z}]&=&\ketbra{m_z}{m'_z}\delta_{0,(m-m')\bmod K}\,.\ea In words, for any observable $A$, the averaging $\mathbb{E}_K$ destroys all the matrix elements in the basis of the eigenvectors of $J_z$, but those such that $(m-m') \bmod K=0$. In particular, when $d\leq K$, $\mathbb{E}_K[A]=\operatorname{diag}_z(A)$ is the diagonal of $A$ in the basis of eigenstates of $J_z$. Since $\mel{m_z}{\sgn(J_x)}{m_z}=0$ for all $m$, it follows that
$Q_K=\frac{1}{2}\one$ and $P_K^{d}=\mathbf{P}_K^{d}=\frac{1}{2}$ for all states, as long as $d\leq K$ i.e.~$j<\frac{K}{2}$. When $d>K$, $\mathbb{E}_K[A]$ has a block-diagonal structure, with $K$ separate subspaces indexed by $\bar{m}\in\{-j,-j+1,...,-j+K-1\}$, each of dimension $\lfloor \frac{j-\bar{m}}{K} \rfloor + 1$. On top of this structure, one uses specific properties of the operator $\sgn(J_x)$. 

All that precedes is detailed in Appendix~\ref{apd:QKd}. In Appendix~\ref{sec:matrix-elements}, we 
use the Wigner matrices that connect the eigenstates of $J_z$ and those of $J_x$ \cite{Sakurai,angular-momentum-in-qm} to obtain the matrix elements of $\sgn(J_x)$ in the basis of the eigenstates of $J_z$. Eventually, the values of $\mathbf{P}_K^{d}$ that we were able to compute analytically are listed in Appendix~\ref{sec:particular-values}. All these values are represented by red dots in Figs.~\ref{fig:wiggly-3} and \ref{fig:wiggly-7}.

In particular, we have obtained the expression of the first nontrivial value \ba
\mathbf{P}_K^{K+1}&=&\,\frac{1}{2}\left[1+2^{-(K-1)}\pmqty{K-1\\\frac{K-1}{2}}\right]\label{probfirst}\,. \ea The numerics strongly support the conjecture that this is the largest $\mathbf{P}_K^{d}$ for given $K$. While we have not been able to prove this conjecture in its generality, we do have the expression for the next several $\mathbf{P}_K^d$, including the next two peaks $\mathbf{P}_K^{3K+1}$ and $\mathbf{P}_K^{5K+1}$, and they are all indeed provably smaller. The eigenvalue $\mathbf{P}_K^{K+1}$ is nondegenerate and its corresponding eigenstate is \ba
\ket{\Psi_K}\,=\, \frac{1}{\sqrt{2}}\bqty{
    \ket{m_z=j} + (-1)^{({K-1})/{2}} \ket{m_z=-j}}
\label{psiopt}
\ea for $j=\frac{K}{2}$. The largest among the $\mathbf{P}_K^{K+1}$ is $\mathbf{P}_3^4=\frac{3}{4}$, obtained when $j=\frac{3}{2}$. Under the conjecture just mentioned, this value is the largest quantum violation of the classical bound.

This protocol is also an entanglement witness, when performed with composite systems. Indeed, if the system is composed of two particles with fixed spins $j_1$ and $j_2$, the state that achieves the maximum violation for measurements of ${J}_x = {J}_x^{(j_1)}\otimes\mathbbm{1}^{(j_2)} + \mathbbm{1}^{(j_1)}\otimes{J}_x^{(j_2)}$ is always entangled when expressed in the subsystems (Appendix~\ref{apd:spin-inseparable}). This can be extended the multipartite case, which is a composite system of many particles, each with a fixed spin. For example, for a chain of $K$ spin-$\frac{1}{2}$ particles, the state that achieves the score $\mathbf{P}_K^{K+1}$ is the maximally entangled $K$-partite Greenberger-Horne-Zeilinger state \begin{equation}
\ket{\Psi_K} = \frac{1}{\sqrt{2}}\bqty\big{\ket{\uparrow}^{\otimes K} + (-1)^{({K-1})/{2}}\ket{\downarrow}^{\otimes K}}\,.
\end{equation}

Most of the structure found for finite-dimensional systems carries over to the quantum harmonic oscillator by replacing $\ket{m_z} \to \ket{n}$, $J_z \to a^\dag a$, and $\pos(J_x) \to \pos(X)$, where
we present these analogies\ba \pos(X) = \int_{0}^\infty\dd{x} \ketbra{x}&=& \frac{1}{2}\bqty{\mathbbm{1}+\sgn(X)}\,.\ea In Appendix~\ref{apd:ho} we present these analogies. In particular, one finds the explicit expression \eqref{eq:sgn-x-mel} for $\mel{n}{\sgn(X)}{n'}$. The plots of Fig.~\ref{fig:wigner-plot} were made by diagonalizing a truncation of this operator. We also show that, in the limit $j\rightarrow\infty$, $\pos(J_x)$ has the same matrix elements as $\pos(X)$. Therefore, as noticed, $\mathbf{P}_K^{d\rightarrow\infty}$ is indeed the same as $\mathbf{P}_K^{\infty}$ described in Sec.~\ref{sec:ho} \footnote{Note that the analogy between a large spin and the harmonic oscillator cannot be based on the approximation $[J_x,J_y]\sim [x,p]$, frequently used for instance in describing collective spins like atomic clocks. That approximation holds only for states such that $\langle J_z\rangle\approx \hbar j$, which is not the case here.}.

We have also obtained bounds for $\mathbf{P}_K^\infty$. On the one hand, in Appendix~\ref{apd:ho} we have the closed-form lower bound
\ba
\mathbf{P}_K^\infty\,\geq\,\mathbf{P}_K^{\text{LB}}&=&\frac{1}{2}\left(1+\sqrt{\frac{1+F_K}{K}}\right)
\label{lowerbound}
\ea with $F_K=\frac{2}{\pi} \sum_{k=1}^{({K-1})/{2}} (-1)^k\arccos[2\sin(\frac{\pi k}{K})-1]$. It can be verified that $\mathbf{P}_K^{\text{LB}}$ is strictly larger than $\mathbf{P}_K^{c}$, for every odd $K$. As for an upper bound, we start by noticing that the block structure of $\EE_K[\pos(X)]$ implies that the eigenvector corresponding to the maximal eigenvalue is a superposition of Fock states differing by integer multiples of $K$. For such vectors, $W(r,\theta+2\pi k/K)=W(r,\theta)$ holds, where $W(r,\theta)\equiv W(r\cos\theta,r\sin\theta)$. In other words, their Wigner function has the same symmetry as the classical probabilities (see Fig.~\ref{fig:wigner-plot}). For all states with this symmetry, \ba P^\infty_K&=&\frac{K+1}{2}\,\mathcal{W}_{+}+\frac{K-1}{2}\,\mathcal{W}_{-}\,=\, \mathbf{P}_K^{c}-\mathcal{W}_{-}\,,\ea where \ban
\mathcal{W}_{\pm}=\int_{\theta\in\Theta_{\pm k}}
    \int_{r\in\mathbb{R}^+}
    r\dd{r}\dd{\theta} W(r,\theta)
\ean are the weights over the $\Theta_{+k}$ and $\Theta_{-k}$ sectors and we have used also the normalization $\mathcal{W}_{+} +\mathcal{W}_{-}=\frac{1}{K}$. The weight of a Wigner function over any pointed sector in phase space is lower bounded as $\mathcal{W}\geq -s_-\approx -0.1559$ \cite{werner-bound}. Thus by setting $\mathcal{W}_{-}=-s_-$ we have
\ba
\mathbf{P}_K^{\infty} \,\leq \, \mathbf{P}_K^{\text{UB}}&=& \mathbf{P}_K^{c} + s_-\,.\label{upperbound}
\ea 
The numerics (see the values reported in Figs.~\ref{fig:wiggly-3} and \ref{fig:wiggly-7}) suggest that $\mathbf{P}_K^{\text{LB}}$ is very close to $\mathbf{P}_K^\infty$. If that is the case, the upper bound \eqref{upperbound} is not tight ($\mathbf{P}_3^{\text{UB}}\approx 0.8226$ and $\mathbf{P}_7^{\text{UB}}\approx 0.7273$); it is nonetheless an improvement, since previously only $\mathbf{P}_K^\infty<1$ was rigorously proved \cite{tsirelson}.

\section{Conclusion}

In summary, we have presented a criterion to certify the nonclassical nature of a single quantum system based on its time evolution, when the latter is a simple precession at frequency $\nu$. Any detection of quantumness must involve incompatible variables and ours is not an exception. A gap with the classical bound can only happen when the precession couples noncommuting variables and is well understood in terms of quasiprobability distributions with negative values \cite{quasiprobability-review}. Our criterion can be tested on many platforms. For discrete systems, the most obvious that come to mind are a proper spin, for instance, a nuclear spin, and Zeeman subspaces in atoms and ions. For continuous variables, one could consider optical, acoustic, or mechanical modes. The dynamics being usually embedded in the system, the challenge lies mainly in the creation of a suitable state and in the detection depending on the platform. 

Detection of nonclassical features has been a prominent topic in recent decades, in the context of emerging quantum technologies. As any violation of the classical bound requires negativity in the Wigner function, our protocol is a negativity witness \cite{negativity-witness}. In terms of usefulness, at a very general level one can say that any of the states detected by our protocol, together with the set of Gaussian operations, is sufficient for universal quantum computation \cite{universal-CV}. More specifically, with the states $\ket{\Psi_K}$ that are optimal for our protocol one can construct an order-$K$ bosonic error-correcting code \cite{bosonic-code} with the logical encoding $\ket{\pm_K} = (\pm 1)^{a^\dag a}\ket{\Psi_K}$ and $Z = e^{i(\pi/K)a^\dag a}$.

Since it does not involve entanglement, our criterion cannot be a ``black box,'' that is, a device-independent certification \cite{RMPnonlocal}. Indeed, two elements of characterization are needed: The dynamics must be a precession at a known pulsation $\omega$ and one must be measuring the same variable ($x$ or $J_x$ in the cases we studied) at any of the probing times $t_k$.

The fact that only one measurement is performed in each round, that of a single dynamical variable, distinguishes our criterion from tests of contextuality \cite{context} and Leggett-Garg-type criteria \cite{LG}, which require the simultaneous or sequential measurement of two or more observables in each round. In particular, our criterion does not utilize the noninvasive measurements required in similar Leggett-Garg tests of harmonic oscillators \cite{LG-coherent,LG-QHM}, which avoids the ``clumsiness'' loophole entirely \cite{LG-clumsy}.

We also highlight that, under the very plausible conjecture that $\max_d\mathbf{P}_K^{d}=\mathbf{P}_K^{K+1}$, the observation of this maximal quantum violation identifies uniquely the dimension and the state, under the assumptions that the dynamics is a precession and that one is measuring a component of angular momentum. This would be a different type of single-particle self-test: Previously reported instances of such tests assumed either a bound on the dimension of the system (see Sec.~9.2 of \cite{selftest-review} for a review) or the properties of a set of observables \cite{selftest}.

Some directions of future research should be noted. First, there is a possible conceptual connection between our protocol and the concept of dynamic nonlocality, so called because the time evolution of some observables depends on the value of the potential at different locations \cite{Popescu}. An example of such observables are modular variables, like $x(t) \bmod x_0$ and $p(t) \bmod p_0$, which have already been considered together with Leggett-Garg inequalities for tests of contextuality \cite{LG-Ramsey}. Modular variables split phase space into periodic cells, analogous to how our protocol splits phase space into sectors (see Fig.~\ref{fig:classical-case}). Further work would be required to investigate if a formal relationship exists between our protocol and that of dynamic nonlocality.

Second, while we have focused on the violation of the classical bound by a \emph{quantum} state, casting this protocol on a general probabilistic theory (GPT) framework is another avenue for future research. It is clear that epistemically restricted theories cannot reproduce the quantum gap, as they recover only a subtheory of quantum mechanics with Gaussian states, which cannot violate the classical bound \cite{GPT-epi}. While there have been recent works on continuous-variable GPT that permit negative quasiprobabilities \cite{GPT-spectrum,GPT-tunnel}, the full set of states allowed under such theories is yet to be characterized.

Beyond these timely concerns, we repeat the core of our surprising result by paraphrasing Tsirelon's title \cite{tsirelson}: Quantumness can be certified simply by asking how often the coordinate of a uniformly precessing system is positive.

\section*{Acknowledgments}

We thank Miguel Navascu\'es for bringing Tsirelson's paper \cite{tsirelson} to our attention and Reinhard Werner for bringing Ref.~\cite{werner-bound} to our attention. We enjoyed discussions with several colleagues at CQT.

We acknowledge financial support from the National Research Foundation and the Ministry of Education, Singapore, under the Research Centres of Excellence program.

\bibliographystyle{apsrev4-2}
\bibliography{refs}

\begin{appendix}

\section{Properties of $Q_K$}
\label{apd:QKd}

\subsection{\label{sec:prop-block-diagonal}Block diagonal structure enforced by $\EE_K$.}
From the main text, $\mathbf{P}^d_K$ is the largest eigenvalue of the observable $Q_K = \EE_K[\pos(J_x)]$. In this section, we will show that $Q_K$, or indeed $\EE_K[A]$ for an arbitrary operator $A = \sum_{m,m'} \mel{m_z}{A}{m_z'} \; \ketbra{m_z}{m_z'}$, is block diagonal. We begin by considering the effect of $\EE_K$ on each basis operator $\ketbra{m_z}{m_z'}$,
\begin{equation}
\begin{aligned}
\EE_K[\ketbra{m_z}{m_z'}] &=
\frac{1}{K}\sum_{k=0}^{K-1} e^{-i (2\pi k/K) J_z} \ketbra{m_z}{m_z'} e^{i(2\pi k/K)J_z} \\
&=
\ketbra{m_z}{m_z'} \frac{1}{K}\sum_{k=0}^{K-1} e^{-i 2\pi k\bqty{(m-m')/K}}. \end{aligned}
\end{equation}
The summation is a sum over the roots of unity, which evaluates to 1 when $K$ is a factor of $m-m'$, and 0 otherwise: $\EE_K[\ketbra{m_z}{m_z'}] = \ketbra{m_z}{m_z'}\delta_{0,(m-m')\bmod K}$.

This means that $\mel{m_z}{\mathbb{E}_K[A]}{m_z'}$ is nonzero only when $m \bmod K = m' \bmod K$, which in turn implies that $\mathbb{E}_K[A]$ has a block diagonal structure, where each unique value of $(m-m') \bmod K \in \{0,1,\cdots,\min(K-1,2j)\}$ defines a separate subspace. Here, $(m-m') \bmod K$ is maximally $\min(K-1,2j)$ as it cannot be greater than or equal to the dimension of the system $d=2j+1$. It is convenient to label the subspaces with $\bar{m} \in \{-j,-j+1,\cdots,\min(-j+K-1,j)\}$, where the $\bar{m}$th block refers to the block containing the state $\ket{m_z = \bar{m}}$.

In summary, we have
\begin{equation}\label{eq:block-diagonal-form}
\begin{aligned}
    \EE_K\bqty{A}
    &= \bigoplus_{\bar{m}=-j}^{\min(j,-j+K-1)}
    \Pi_K^{(\bar{m})} A \Pi_K^{(\bar{m})},
\end{aligned}
\end{equation}
where $\Pi_K^{(\bar{m})}$ is a projection into the $\bar{m}$th subspace,
\begin{equation}
    \Pi_K^{(\bar{m})} \equiv \sum_{k=0}^{\lfloor({j-\bar{m}})/{K}\rfloor} \ketbra{m_z=\bar{m}+kK}.
\end{equation}
The upper limit comes from the requirement that $m_z = \bar{m}+kK \leq j$ and it follows that the dimension of the $\bar{m}$th subspace is $\lfloor\frac{j-\bar{m}}{K}\rfloor + 1$.

Notice that when $d \leq K$, $\lfloor\frac{j-\bar{m}}{K}\rfloor \leq \lfloor\frac{d-1}{K}\rfloor = 0$, so the dimension of every subspace is 1. This means that
\begin{equation}\label{eq:block-diagonal-diagonal}
    \EE_K\bqty{A} = \operatorname{diag}_z(A)\qquad\text{when}\;d \leq K.
\end{equation}

For later convenience, we define further splits of each subspace $\Pi_K^{(\bar{m})}$ into even and odd subspaces
\ba \Pi_K^{(\bar{m})} = \Pi_K^{(+\bar{m})} + \Pi_K^{(-\bar{m})}\,,\ea where
\begin{equation}
\begin{aligned}
    \Pi_K^{(+\bar{m})} &\equiv \sum_{k=0}^{\lfloor \pqty{j-\bar{m}}/{2K} \rfloor} \ketbra{m_z = \bar{m}+2kK},\\
    \Pi_K^{(-\bar{m})} &\equiv \hspace{-.75em} \sum_{k=0}^{\lfloor \pqty{j-\bar{m}-K}/{2K} \rfloor} \hspace{-.75em} \ketbra{m_z = \bar{m}+(2k+1)K}{\bar{m}+(2k+1)K}.
\end{aligned}\label{twosubspaces}
\end{equation}

\subsection{\label{sec:prop-sgn-Jx}Properties of $\sgn(J_x)$}
The observable $\pos(J_x)$ can be rewritten as
\ba\pos(J_x) = \frac{1}{2}\bqty{\mathbbm{1} + \sgn(J_x)}\,,\ea where \ba\sgn(J_x) = \sum_{m=-j}^j \sgn(m)\ketbra{m_x}\,.\ea We do so because there are several desirable properties of $\sgn(J_x)$ that will simplify later calculations. Clearly, upon a $\pi$ rotation around the $z$ axis, we have
\begin{equation}\label{eq:parity}
    e^{-i\pi J_z}\sgn(J_x)e^{i\pi J_z}
    = -\sgn(J_x).
\end{equation}
From this, we can infer two properties.
\begin{enumerate}[label=\emph{Property \arabic*}.,ref=\arabic*,wide=0pt]
    \item\label{prop-parity} When $m'-m$ is even, $\mel{m_z}{\sgn(J_x)}{m_z'}=0$. Indeed, \begin{equation}
\begin{aligned}
    &\mel{m_z}{\sgn(J_x)}{m_z'} \\
    &\quad{}={} -\bra{m_z} e^{-i\pi J_z}\sgn(J_x)e^{i\pi J_z} \ket{m_z'} \\
    &\quad{}={} -(-1)^{m'-m}\bra{m_z}\sgn(J_x)\ket{m_z'},
\end{aligned}
\end{equation}
    \item\label{prop-dual-eigenvalues} If $\lambda$ is an eigenvalue of $\EE_K[\sgn(J_x)]$, so is $-\lambda$. Indeed, assume that $\ket{\lambda}$ is an eigenvector of $\EE_K[\sgn(J_x)]$ with eigenvalue $\lambda$. Then, 
\begin{equation}
\begin{aligned}
    &\EE_K[\sgn(J_x)]\pqty{e^{i\pi J_z}\ket{\lambda}}\\
    &\quad{}={} e^{i\pi J_z}\EE_K\bqty{e^{-i\pi J_z}\sgn(J_x)e^{i\pi J_z}}\ket{\lambda}\\
    &\quad{}={} -\lambda\pqty{e^{i\pi J_z}\ket{\lambda}},
\end{aligned}
\end{equation} where $e^{i\pi J_z}$ can be dragged into $\EE_K[\cdots]$ in the second step since it commutes with $e^{i(2\pi k/K)J_z}$. Therefore, $e^{i\pi J_z}\ket{\lambda}$ is an eigenvector of $\sgn(J_x)$ with eigenvalue $-\lambda$.
\end{enumerate}

\subsection{\label{sec:even-odd-subspace}Even and odd subspace of $(\EE_K[\sgn(J_x)])^2$}
Section \ref{sec:prop-block-diagonal} tells us that $\EE_K[\sgn(J_x)]$ takes on a block-diagonal form, where $\Pi_K^{(\bar{m})}$ projects onto the $\bar{m}$th subspace. 
Due to property \ref{prop-parity} of $\sgn(J_x)$, $\Pi_K^{(\pm\bar{m})}\sgn(J_x)\Pi_K^{(\pm\bar{m})} = 0$. Therefore, using \eqref{twosubspaces},
\begin{equation}\label{eq:even-odd-separation}
\begin{aligned}
    &\bqty{\Pi_K^{(\bar{m})}\sgn(J_x)\Pi_K^{(\bar{m})}}^2\\
    &= \bqty{\Pi_K^{(+\bar{m})}\sgn(J_x)\Pi_K^{(-\bar{m})} +
    \Pi_K^{(-\bar{m})}\sgn(J_x)\Pi_K^{(+\bar{m})}}^2 \\
    &= \Pi_K^{(+\bar{m})}\sgn(J_x)\Pi_K^{(-\bar{m})}\sgn(J_x)\Pi_K^{(+\bar{m})} \\
    &\qquad{}+{}\Pi_K^{(-\bar{m})}\sgn(J_x)\Pi_K^{(+\bar{m})}\sgn(J_x)\Pi_K^{(-\bar{m})},
\end{aligned}
\end{equation}
that is, each block of $\{\EE_K[\sgn(J_x)]\}^2$ is itself made up of a separate even and odd block labeled by $+\bar{m}$ and $-\bar{m}$. Furthermore, since the two terms in Eq.~\eqref{eq:even-odd-separation} are of the form $AA^\dag$ and $A^\dag A$, respectively, they share the same nonzero eigenvalues.

Finally, if $\lambda_2$ is the maximum eigenvalue of $\{\EE_K[\sgn(J_x)]\}^2$, then $\lambda=\sqrt{\lambda_2}$ is the maximum eigenvalue of $\EE_K[\sgn(J_x)]$. There is no ambiguity with the signs, as the nonzero eigenvalues of $\EE_K[\sgn(J_x)]$ must come in $(\lambda,-\lambda)$ pairs due to property \ref{prop-dual-eigenvalues} of $\sgn(J_x)$.

What we conclude through this line of reasoning is that we can find the maximum eigenvalue $\lambda$ over just the odd (or equivalently, even) blocks,
\begin{equation}
\begin{aligned}
    \lambda_2 = 
    \hspace{-0.75em}\max_{
        \begin{array}{c}
            \bar{m},\ket{\psi},\\[-.5ex]
            \scriptstyle\braket{\psi}=1
        \end{array}
    }\hspace{-0.75em} \bra{\psi}\Pi_K^{(-\bar{m})}\sgn(J_x)\Pi_K^{(+\bar{m})}\sgn(J_x)\Pi_K^{(-\bar{m})}\ket{\psi},
\end{aligned}
\end{equation}
to obtain $\mathbf{P}_K^d = \frac{1}{2}(1+\sqrt{\lambda_2})$.

\section{\label{sec:matrix-elements}Matrix Elements of $\sgn(J_x)$}
Throughout this appendix we will only consider $m'$ and $m$ such that $(m'-m)\bmod 2 \neq 0$, as the matrix elements of $\sgn(J_x)$ are zero otherwise. An explicit expression for $\mel{m_z}{\sgn(J_x)}{m_z'}$ is given in Eq.~\eqref{eq:sgn-Jx-mel}, the derivation of which will occupy the rest of the appendix:
\begin{widetext}
\begin{equation}\label{eq:sgn-Jx-mel}
\begin{aligned}
    \mel{m_z}{\sgn(J_x)}{m_z'}
    &=
    \frac{(-1)^{\pqty{m'-m-1}/{2}}2^{-(2j-1)}}{m'-m}
    \sqrt{
        \pmqty{
            2\lfloor \frac{j+m}{2} \rfloor \\
             \lfloor \frac{j+m}{2} \rfloor
        }
        \pmqty{
            2\lfloor \frac{j-m}{2} \rfloor \\
             \lfloor \frac{j-m}{2} \rfloor
        }
        \pmqty{
            2\lfloor \frac{j+m'}{2} \rfloor \\
             \lfloor \frac{j+m'}{2} \rfloor
        }
        \pmqty{
            2\lfloor \frac{j-m'}{2} \rfloor \\
             \lfloor \frac{j-m'}{2} \rfloor
        }
    }\\
    &\qquad\qquad{}\times{}\sqrt{
        (j+m)^{(j+m)\bmod 2}
        (j+m')^{(j+m')\bmod 2}
        (j-m)^{(j-m)\bmod 2}
        (j-m')^{(j-m')\bmod 2}
    }
\end{aligned}
\end{equation}
\end{widetext}
A small simplification of $\mel{m_z}{\sgn(J_x)}{m_z'}$ is provided before continuing with the rest of the derivation. Here $\Delta^j_{m,m'} \equiv \braket{m_z}{m_x'} = \mel{m_z}{e^{-i(\pi/2)J_y}}{m_z'}$ is an element of the Wigner $d$ matrix with some known symmetries and values \cite{Sakurai}:
\begin{align}\label{eq:Wigner-d-sym}
    \Delta^j_{m,m'} &= (-1)^{m-m'}\Delta^j_{m',m} = (-1)^{j+m}\Delta^j_{m,-m'}, \\ \label{eq:Wigner-d-value-j}
    \Delta^j_{m,j} &= 2^{-j}\sqrt{\pmqty{2j\\j-m}}, \\ \label{eq:Wigner-d-value-0}
    \Delta^j_{m,0} &= \begin{cases}
        0\qquad\qquad\text{for $j+m$ odd; otherwise}, \\
        (-1)^{\pqty{j+m}/{2}}2^{-j}\sqrt{
            \pmqty{j+m\\\frac{j+m}{2}}
            \pmqty{j-m\\\frac{j-m}{2}}
        }.
    \end{cases}
\end{align}
Some recursion formulas are also known, including \cite{angular-momentum-in-qm}
\begin{equation}\label{eq:Wigner-d-recursive}
\begin{aligned}
    \Delta^j_{m,m'}
    &= \frac{1}{\sqrt{2(j-m')}}\bigg(
        \sqrt{j-m}\Delta^{j-\frac{1}{2}}_{m+\frac{1}{2},m'+\frac{1}{2}} \\
        &\hspace{8.5em}{}-{} \sqrt{j+m}\Delta^{j-\frac{1}{2}}_{m-\frac{1}{2},m'+\frac{1}{2}}
    \bigg).
\end{aligned}
\end{equation}
Then, rewriting the matrix elements of $\sgn(J_x)$ in terms of $\Delta^j_{m,m'}$ and using Eq.~\eqref{eq:Wigner-d-sym},
\begin{equation}
\begin{aligned}
\mel{m_z}{\sgn(J_x)}{m_z'}
&= \sum_{m''=-j}^{j} \sgn(m'') \Delta^j_{m,m''}\Delta^j_{m',m''}\\
&= 2 \sum_{m''=m_0}^{j} \Delta^j_{m,m''}\Delta^j_{m',m''},
\end{aligned}
\end{equation}
where $m_0 = 1/2$ if $d$ is even and $m_0 = 1$ if $d$ is odd.

\subsection{\label{sec:sgn-Jx-mel-edge}Edge Cases $\mel{m_z=-j}{\sgn(J_x)}{m_z=-j+k}$}
We begin the derivation with edge cases of the form $\mel{m_z=-j}{\sgn(J_x)}{m_z = -j+k}$, where $k > 0$ is odd:
\begin{equation}\label{eq:sgn-Jx-mel-edge-step-1}
\begin{aligned}
    &\mel{m_z=-j}{\sgn(J_x)}{m_z=-j+k} \\
    &\quad{}={} 2\Delta^j_{-j,j}\Delta^j_{-j+k,j} + 2\sum_{m=m_0}^{j-1} \Delta^j_{-j,m}\Delta^j_{-j+k,m}.
\end{aligned}
\end{equation}
The first term can be resolved with Eq.~\eqref{eq:Wigner-d-value-j},
\begin{equation}\label{eq:sgn-Jx-mel-edge-first}
\begin{aligned}
    2\Delta^j_{j,-j}\Delta^j_{j,-j+k}
    &= 2\cdot2^{-2j} \sqrt{\frac{(2j)!}{(2j-k)!k!}} \\
    &= \sqrt{\frac{2j}{k}}\Delta^{j-\frac{1}{2}}_{j-\frac{1}{2},j-\frac{1}{2}}
    \Delta^{j-\frac{1}{2}}_{j-\frac{1}{2},j-k+\frac{1}{2}},
\end{aligned}
\end{equation}
where we have rewritten it in a form for later convenience. Meanwhile, with the recursive relation from Eq.~\eqref{eq:Wigner-d-recursive}, the second term in Eq.~\eqref{eq:sgn-Jx-mel-edge-step-1} is
\begin{align}
    &2\sum_{m=m_0}^{j-1} \Delta^j_{-j,m}\Delta^j_{-j+k,m}
    =  (-1)^k2\sum_{m=m_0}^{j-1} \Delta^j_{m,j}\Delta^j_{m,j-k} \nonumber\\
    &= -2\sum_{m_0+1}^j \Delta^j_{m-1,j}
    \sqrt{\frac{j-m+1}{2k}} \Delta^{j-\frac{1}{2}}_{m-\frac{1}{2},j-k+\frac{1}{2}} \nonumber\\
    &\qquad{}+{} 2\sum_{m_0}^{j-1} \Delta^j_{m,j}
    \sqrt{\frac{j+m}{2k}} \Delta^{j-\frac{1}{2}}_{m-\frac{1}{2},j-k+\frac{1}{2}} \label{eq:sgn-Jx-mel-edge-second}\\
    &= - 2\sqrt{\frac{1}{2k}}
    \Delta^j_{j-1,j}\Delta^{j-\frac{1}{2}}_{j-\frac{1}{2},j-k+\frac{1}{2}}
    + 2\sqrt{\frac{j+m_0}{2k}}\Delta^j_{m_0,j}\Delta^{j-\frac{1}{2}}_{m_0-\frac{1}{2},j-k+\frac{1}{2}}\nonumber\\
    &\qquad{}-{} 2\sum_{m_0+1}^{j-1} \underbrace{\pqty{
        \Delta^j_{m-1,j}\sqrt{\tfrac{j-m+1}{2k}}
        - \Delta^j_{m,j}\sqrt{\tfrac{j+m}{2k}} 
    }}_{=\,0,\;\text{using equation}~\eqref{eq:Wigner-d-value-j}}
    \Delta^{j-\frac{1}{2}}_{m-\frac{1}{2},j-k+\frac{1}{2}}\nonumber\\
    &= - 2\sqrt{\frac{1}{2k}}
    \pqty{2^{-j}\sqrt{\frac{(2j)!}{(2j-1)!2!}} }
    \Delta^{j-\frac{1}{2}}_{j-\frac{1}{2},j-k+\frac{1}{2}}\nonumber\\
    &\qquad{}+{} 2\sqrt{\frac{j+m_0}{2k}}
    \pqty{2^{-j}\sqrt{\frac{(2j)!}{(j-m_0)!(j+m_0)!}} }
    \Delta^{j-\frac{1}{2}}_{m_0-\frac{1}{2},j-k+\frac{1}{2}}\nonumber\\
    &= -\sqrt{\frac{2j}{k}}\Delta^{j-\frac{1}{2}}_{j-\frac{1}{2},j-\frac{1}{2}}
        \Delta^{j-\frac{1}{2}}_{j-\frac{1}{2},j-k+\frac{1}{2}}\nonumber\\
        &\qquad{}+{} \sqrt{\frac{2j}{k}}\Delta^{j-\frac{1}{2}}_{m_0-\frac{1}{2},j-\frac{1}{2}}
        \Delta^{j-\frac{1}{2}}_{m_0-\frac{1}{2},j-k+\frac{1}{2}} \nonumber
\end{align}
Recognizing the first term as the negative of that in Eq.~\eqref{eq:sgn-Jx-mel-edge-first}, Eq.~\eqref{eq:sgn-Jx-mel-edge-step-1} is therefore
\begin{equation}\label{eq:sgn-Jx-mel-edge-step-2}
\begin{aligned}
    &\mel{m_z=-j}{\sgn(J_x)}{m_z=-j+k} \\
    &\quad{}={} \sqrt{\frac{2j}{k}}\Delta^{j-\frac{1}{2}}_{m_0-\frac{1}{2},j-\frac{1}{2}}
        \Delta^{j-\frac{1}{2}}_{m_0-\frac{1}{2},j-k+\frac{1}{2}}.
\end{aligned}
\end{equation}
If $d$ is even, the terms are already in the form given in Eq.~\eqref{eq:Wigner-d-value-j}. If $d$ is odd, the recursive relation from Eq.~\eqref{eq:Wigner-d-recursive} can be applied once more. In both cases,
\begin{widetext}
\begin{equation}\label{eq:sgn-Jx-mel-edge}
\mel{m_z=-j}{\sgn(J_x)}{m_z=-j+k}
= (-1)^{\pqty{k-1}/{2}} 2^{-(2j-1)}\sqrt{
        \pqty{\frac{2j}{k}-\pqty{d\bmod{2}} }
        \pmqty{
            2\lfloor j \rfloor \\
            \lfloor j \rfloor
        }
        \pmqty{
            2\lfloor j-\frac{k}{2} \rfloor \\
            \lfloor j-\frac{k}{2} \rfloor
        }
        \pmqty{
            k-1 \\
            \frac{k-1}{2}
        }
    }.
\end{equation}
\end{widetext}

\subsection{Commutator of $\sgn(J_x)$ and $J_-$}
In this section we work out how $\sgn(J_x)$ commutes with the lowering operator $J_- = J_x - iJ_y$ to find the rest of the matrix elements. First, we have
\begin{align}
    &\comm{J_-}{\sgn(J_x)} 
    = \comm{J_x-iJ_y}{\sgn(J_x)} \nonumber\\
    &\quad{}={} -i \comm{J_y}{e^{-i(\pi/2)J_y}\sgn(J_z)e^{i(\pi/2)J_y}}\\
    &\quad{}={} \frac{1}{2} e^{-i(\pi/2)J_y}\pqty\Big{
        - \comm{J_+}{\sgn(J_z)} + \comm{J_-}{\sgn(J_z)}
    } e^{i(\pi/2)J_y}.\nonumber
\end{align}
For any $\abs{m} > 1$, $\sgn(m\pm1)=\sgn(m)$. As such, the commutators would only depend on the states $\ket{m_z}$ with $\abs{m} \leq 1$. The action of the commutators on these states is easily found. For $d$ even,
\begin{align}
    &\comm{J_\mp}{\sgn(J_z)}\ket{\pm\tfrac{1}{2}_z} \nonumber\\
    &\quad{}={} \pm J_\mp\ket{\pm\tfrac{1}{2}_z} - \sgn(J_z)J_\mp\ket{\pm\tfrac{1}{2}_z}\\
    &\quad{}={} \pm (2j+1)\ket{\mp\tfrac{1}{2}_z},\nonumber
\end{align}
and for $d$ odd,
\begin{align}
    \comm{J_\mp}{\sgn(J_z)}\ket{\pm1_z}
    &= \pm\sqrt{j\pqty{j+1}}\ket{\mp1_z},\\
    \comm{J_\mp}{\sgn(J_z)}\ket{0_z}
    &= -\sqrt{j(j+1)}\ket{\mp1_z}.\nonumber
\end{align}
Therefore, the commutator of $\sgn(J_x)$ and $J_-$ is
\begin{equation}\label{eq:sgn-Jx-commutator}
\begin{aligned}
    &\comm{J_-}{\sgn(J_x)} \\
    &\;{}={} \begin{cases}
        (j+\frac{1}{2})\pqty{
            \ketbra{-\tfrac{1}{2}_x}{\tfrac{1}{2}_x} +
            \ketbra{\tfrac{1}{2}_x}{-\tfrac{1}{2}_x}
        } & \text{for $d$ even} \\
        \begin{array}{l}
            \sqrt{j(j+\frac{1}{2})}\Big(
                \frac{1}{\sqrt{2}}\pqty\big{\ket{1_x}+\ket{-1_x}}\bra{0_x}\\
             \quad {}+{}
                \ket{0_x}\frac{1}{\sqrt{2}}\pqty\big{\bra{1_x}+\bra{-1_x}}
            \Big)
        \end{array} & \text{for $d$ odd}.
    \end{cases}
\end{aligned}
\end{equation}

\subsection{\label{sec:sgn-Jx-mel-other-even}Other matrix elements of $\sgn(J_x)$ for $d$ even}
This section completes the derivation of the matrix elements of $\sgn(J_x)$ for $d$ even, focusing on values of $\mel{m_z}{\sgn(J_x)}{m_z'}$ where $m' > m$, as the converse case can be found from $\sgn(J_x)^\dag = \sgn(J_x)$. By defining $l \equiv m+j$ and $k \equiv m'-m$, these matrix element can be written as $\mel{m_z=-j+l}{\sgn(J_x)}{m_z=-j+l+k}$, which is in a form similar to the edge cases in Appendix~\ref{sec:sgn-Jx-mel-edge}, but with an offset of $l$.

Our approach is to set up a recursive relation using the commutator found in Eq.~\eqref{eq:sgn-Jx-commutator} and using the action of the raising operator $J_+\ket{m_z=j+l-1} = \sqrt{l(2j-l+1)}\ket{m_z=-j+l}$. Then
\begin{widetext}
\begin{align}
    &\mel{m_z=-j+l}{\sgn(J_x)}{m_z=-j+k+l}\nonumber\\\label{eq:sgn-Jx-recursive}
    &= \quad\frac{1}{\sqrt{l(2j-l+1)}}\mel{m_z=-j+l-1}{J_-\sgn(J_x)}{m_z=-j+k+l} \\
    &= \quad\frac{1}{\sqrt{l(2j-l+1)}}\mel{m_z=-j+l-1}{\pqty\big{\sgn(J_x) J_- + \comm{J_-}{\sgn(J_x)}}}{m_z=-j+k+l} \nonumber\\
    &= \quad\sqrt{\frac{(l+k)(2j-l+1-k)}{l(2j-l+1)}}\mel{m_z=-j+l-1}{\sgn(J_x)}{m_z=-j+k+l-1}
    + \frac{(-1)^{l+1}(2j+1)}{\sqrt{l(2j-l+1)}} \Delta^j_{-j+l-1,-\frac{1}{2}}
    \Delta^j_{-j+l+1,-\frac{1}{2}}.\nonumber
\end{align}
\end{widetext}
This sets up a recursive relation between the matrix element offset by $l$ and $l-1$, as desired. Meanwhile, the second term can be found using Eqs.~\eqref{eq:Wigner-d-value-j} and \eqref{eq:Wigner-d-recursive},
\begin{widetext}
\begin{equation}\label{eq:sgn-Jx-recursive-term}
\begin{aligned}
&\frac{(-1)^{l+1}(2j+1)}{\sqrt{l(2j-l+1)}} \Delta^j_{-j+l-1,-\frac{1}{2}}
    \Delta^j_{-j+l+l,-\frac{1}{2}} \\
&\quad{}={} \frac{(-1)^{\pqty{k-1}/{2}}2^{-(2j-1)}}{\sqrt{l(2j-l+1)}}
\times \begin{cases}
    \sqrt{
        (2j-l+1)(2j-k+l)
        \pmqty{l-1\\\frac{l-1}{2}}
        \pmqty{2j-l\\\frac{2j-l}{2}}
        \pmqty{k+l\\\frac{k+l}{2}}
        \pmqty{2j-1-(k+l)\\\frac{2j-1-(k+l)}{2}}
    } & \text{$l$ odd},\\[3ex]
    -\sqrt{
        (l-1)(k+l)
        \pmqty{l-2\\\frac{l-2}{2}}
        \pmqty{2j-l+1\\\frac{2j-l+1}{2}}
        \pmqty{k+l-1\\\frac{k+l-1}{2}}
        \pmqty{2j-(k+l)\\\frac{2j-(k+l)}{2}}
    } & \text{$l$ even}.
\end{cases}
\end{aligned}
\end{equation}
\end{widetext}
Finally, we guess that the matrix element is of the form
\begin{widetext}
\begin{equation}\label{eq:sgn-Jx-mel-final}
\begin{aligned}
&\mel{m_z=-j+l}{\sgn(J_x)}{m_z=-j+k+l}\\
&\quad{}={} \frac{(-1)^{\pqty{k-1}/{2}}2^{-(2j-1)}}{k}
\times\begin{cases}
    \sqrt{
        l(2j-(k+l))
        \pmqty{l-1\\\frac{l-1}{2}}
        \pmqty{2j-l\\\frac{2j-l}{2}}
        \pmqty{k+l\\\frac{k+l}{2}}
        \pmqty{2j-1-(k+l)\\\frac{2j-1-(k+l)}{2}}
    } & \text{$l$ odd},\\[3ex]
    \sqrt{
        (2j-l)(k+l)
        \pmqty{l\\\frac{l}{2}}
        \pmqty{2j-1-l\\\frac{2j-1-l}{2}}
        \pmqty{k+l-1\\\frac{k+l-1}{2}}
        \pmqty{2j-(k+l)\\\frac{2j-k-l}{2}}
    } & \text{$l$ even}.
\end{cases}
\end{aligned}
\end{equation}
\end{widetext}
If Eq.~\eqref{eq:sgn-Jx-mel-final} is true for $l$, due to the recursive relation given in Eq.~\eqref{eq:sgn-Jx-recursive}, it will also be true for $l+1$. Furthermore, since Eq.~\eqref{eq:sgn-Jx-mel-final} with $l=0$ is the same as Eq.~\eqref{eq:sgn-Jx-mel-edge}, it is true for $l=0$. Therefore, by induction, it holds for all values of $l$.

Finally, reinstating $m=-j+l$ and $m'=-j+k+l$ provides us with Eq.~\eqref{eq:sgn-Jx-mel} for $d$ even.

\subsection{\label{sec:sgn-Jx-mel-other-odd}Other matrix elements of $\sgn(J_x)$ for $d$ odd}
The steps required for finding the other matrix elements of $\sgn(J_x)$ for $d$ odd are the same as those given in the preceding section, but with the corresponding commutation relation from Eq.~\eqref{eq:sgn-Jx-commutator}. We find the recursive relation to be
\begin{widetext}
\begin{equation}\label{eq:sgn-Jx-recursive-even}
\begin{aligned}
    &\mel{m_z=-j+l}{\sgn(J_x)}{m_z=-j+k+l}\\
    &= \sqrt{\frac{(k+l)(2j-(k+(l-1))}{l(2j-(l-1))}}\mel{m_z=-j+l-1}{\sgn(J_x)}{m_z=-j+k+l-1}\\
    &\quad
    {}-{} \frac{(-1)^{\pqty{k-1}/{2}}2^{-(2j-1)}[2j-(l-1)-(k+l)]}{2\sqrt{l[2j-(l-1)]}}
    \times\begin{cases}
    \sqrt{
        \pmqty{l-1\\\frac{l-1}{2}}
        \pmqty{2j-(l-1)\\\frac{2j-(l-1)}{2}}
        \pmqty{l+k\\\frac{l+k}{2}}
        \pmqty{2j-(l+k)\\\frac{2j-(l+k)}{2}}
    }
    &\text{$l$ odd},\\
    0&\text{$l$ even}.
    \end{cases}
\end{aligned}
\end{equation}
\end{widetext}
Meanwhile, the corresponding guess is
\begin{widetext}
\begin{equation}\label{eq:sgn-Jx-mel-final-even}
\begin{aligned}
    &\mel{m_z=-j+l}{\sgn(J_x)}{m_z=-j+l+k}\\
    &=
    \frac{(-1)^{\frac{k-1}{2}}2^{-(2j-1)}}{k}
    \times\begin{cases}
    \sqrt{
        l(2j-l)
        \pmqty{l-1\\\frac{l-1}{2}}
        \pmqty{2j-l-1\\\frac{2j-l-1}{2}}
        \pmqty{l+k\\\frac{l+k}{2}}
        \pmqty{2j-(l+k)\\\frac{2j-(l+k)}{2}}
    }
    &\text{$l$ odd},\\
    \sqrt{
        (k+l)[2j-(k+l)]
        \pmqty{l\\\frac{l}{2}}
        \pmqty{2j-l\\\frac{2j-l}{2}}
        \pmqty{l+k-1\\\frac{l+k-1}{2}}
        \pmqty{2j-(l+k)-1\\\frac{2j-(l+k)-1}{2}}
    }
    &\text{$l$ even}.
    \end{cases}
\end{aligned}
\end{equation}
\end{widetext}
Once again, the recursive relation in Eq.~\eqref{eq:sgn-Jx-recursive-even} means that \eqref{eq:sgn-Jx-mel-final-even} being true for $l$ implies it being true for $l+1$. The consistency of Eqs.~\eqref{eq:sgn-Jx-mel-edge} and \eqref{eq:sgn-Jx-mel-final-even} with $l=0$ means that it holds for all values of $l$, providing us with Eq.~\eqref{eq:sgn-Jx-mel} for $d$ odd.

\section{\label{sec:particular-values}Particular Values of $\mathbf{P}_K^d$}
\subsection{For $K$ even}
When $K = 2p$, where $p$ is a positive integer,
\begin{align}
    &e^{-i\pi J_z}\EE_K[\sgn(J_x)] e^{-i\pi J_z} \nonumber\\
    &= \frac{1}{2p}\sum_{k=0}^{p-1}\bigg[
        e^{-i \frac{2\pi}{2p}k J_z}\sgn(J_x) e^{i \frac{2\pi}{2p} k J_z} \\
        &\hspace{6.5em}{}+{}\nonumber
        e^{-i \frac{2\pi}{2p}(p+k) J_z}\sgn(J_x) e^{i \frac{2\pi}{2p}(p+k) J_z}
    \bigg] \\
    &= \frac{1}{2p}\sum_{k=0}^{p-1}
    e^{-i \frac{2\pi}{2p}k J_z}\bqty{
    \sgn(J_x) + e^{-i\pi J_z} \sgn(J_x) e^{i\pi J_z}
    }e^{i \frac{2\pi}{2p}k J_z}\nonumber \\
    &= 0\nonumber,
\end{align}
where we have used Eq.~\eqref{eq:parity} in the last step. Therefore, when $K$ is even, $Q_K = \frac{1}{2}\mathbbm{1}$ and $\mathbf{P}_K^d = \frac{1}{2}$.

\subsection{For $d \leq K$}
When $d \leq K$, from equation \eqref{eq:block-diagonal-diagonal}, $\EE_K[\pos(J_x)] = \Bqty{\mathbbm{1} + \operatorname{diag}_z\bqty{\sgn(J_x)}}/2$. However, property \ref{prop-parity} of $\sgn(J_x)$ means that the diagonal elements are $\mel{m_z}{\sgn(J_x)}{m_z} = 0$. Therefore, $Q_K = \frac{1}{2}\mathbbm{1}$ and $\mathbf{P}_K^d = \frac{1}{2}$.

\subsection{For $K < d \leq 2K$}
When $K < d \leq 2K$, the $\bar{m}$th block is two dimensional for $\bar{m} > j - K$ and one-dimensional for $\bar{m} \leq j - K$. Given property \ref{prop-parity}, the only nonzero elements of $\EE_K[\sgn(J_x)]$ are the off-diagonal terms $\mel{\bar{m}}{\sgn(J_x)}{\bar{m}+K} = \mel{\bar{m}+K}{\sgn(J_x)}{\bar{m}}$, so each two-dimensional block is of the form
\begin{equation}
    \Pi_K^{(\bar{m})}\sgn(J_x)\Pi_K^{(\bar{m})} \widehat{=}
    \mel{\bar{m}}{\sgn(J_x)}{\bar{m}+K}\pmqty{0&1\\1&0},
\end{equation}
which can be easily solved. Using the explicit expressions of the matrix elements in Appendix~\ref{sec:matrix-elements}, the maximum eigenvalue and corresponding eigenvector of $Q_K$ are
\begin{equation}
\begin{aligned}
    \mathbf{P}_K^{d}
    &= \frac{1}{2}\Bigg( 1 + 2^{2j-1}\sqrt{
        \pmqty{K-1\\\frac{K-1}{2}}
        \pqty{\frac{2j}{K} - (d \bmod 2)}
    } \\
    &\hspace{7.5em}{}\times{}\sqrt{
        \pmqty{2\lfloor j \rfloor\\\lfloor j \rfloor}
        \pmqty{2\lfloor j - \frac{K}{2} \rfloor\\\lfloor j - \frac{K}{2} \rfloor}
    }\Bigg), \\
    \ket{\Psi_K} &= \frac{1}{\sqrt{2}}\bqty{
        \ket{m_z = -j+K} + (-1)^{\pqty{K-1}/{2}}\ket{m_z = -j} 
    }.
\end{aligned}
\end{equation}

\subsection{For $2K < d \leq 3K$}
When $2K < d \leq 3K$, the $\bar{m}$th block is three dimensional for $\bar{m} > j - K$ and two dimensional for $\bar{m} \leq j - K$. The two-dimensional blocks are solved in the same manner as before, while the three-dimensional blocks are of the form
\begin{align}
    &\Pi_K^{(\bar{m})}\sgn(J_x)\Pi_K^{(\bar{m})} {}\widehat{=}{}
    \pmqty{0&y_1&0\\y_1&0&y_2\\0&y_2&0},\\[1.5ex]
    &\quad\begin{array}{rl}
    \text{where}\quad y_1 &= \mel{\bar{m}}{\sgn(J_x)}{\bar{m}+K}, \\[.75ex]
    y_2 &= \mel{\bar{m}+K}{\sgn(J_x)}{\bar{m}+2K}.
    \end{array}\nonumber
\end{align}
This can also be easily solved to find
\begin{equation}
\begin{aligned}
    \mathbf{P}_K^{d}
    &= \frac{1}{2}\Bigg[ 1 + \frac{2^{2j-1}}{\cos(\phi)}\sqrt{
        \pmqty{K-1\\\frac{K-1}{2}}
        \pqty{\frac{2j}{K} - (d \bmod 2)}
    } \\
    &\hspace{7.5em}{}\times{}\sqrt{
        \pmqty{2\lfloor j \rfloor\\\lfloor j \rfloor}
        \pmqty{2\lfloor j - \frac{K}{2} \rfloor\\\lfloor j - \frac{K}{2} \rfloor}
    }\Bigg], \\
    \ket{\Psi_K} &= \frac{1}{\sqrt{2}}\Big[
        \cos(\phi)\ket{m_z = -j+2K} \\
        &\qquad\qquad{}+{} (-1)^{\pqty{K-1}/{2}}\ket{m_z = -j+K} \\
        &\qquad\qquad{}+{} \sin(\phi)\ket{m_z = -j}
    \Big],
\end{aligned}
\end{equation}
where
\begin{equation*}
    \phi = \arctan( \sqrt{
        \frac{
            \bqty{j-k\times(d \bmod 2)}
            \pmqty{2\lfloor j \rfloor\\\lfloor j \rfloor}
        }{
            \pqty{j-k}
            \pmqty{2\lfloor j-K \rfloor\\\lfloor j-K \rfloor}
            \pmqty{2K\\K}
        }
    } ).
\end{equation*}

\subsection{For $3K < d \leq 5K$}
For $3K < d \leq 5K$, we can use what we found in Appendix~\ref{sec:even-odd-subspace} to simplify each four- and five-dimensional blocks into two-dimensional ones. In particular, the $-\bar{m}$ subspace of $\{\EE_K[\sgn(J_x)]\}^2$ is
\begin{align}
&\Pi_K^{(-\bar{m})}\sgn(J_x)\Pi_K^{(+\bar{m})}\sgn(J_x)\Pi_K^{(-\bar{m})}
{}\widehat{=}{}\pmqty{
y_1 & y_2 \\
y_2 & y_3
},\\[1.5ex]
&\begin{array}{rl}
\text{where}\quad y_1 &= \sum_{k=0}^{1,2} \abs{\mel{\bar{m}+K}{\sgn(J_x)}{\bar{m}+2kK}}^2, \\[.75ex]
y_2 &= \sum_{k=0}^{1,2} \mel{\bar{m}+K}{\sgn(J_x)}{\bar{m}+2kK}\\
    &\qquad\qquad{}\times{}\mel{\bar{m}+3K}{\sgn(J_x)}{\bar{m}+2kK}, \\[.75ex]
y_3 &= \sum_{k=0}^{1,2} \abs{\mel{\bar{m}+3K}{\sgn(J_x)}{\bar{m}+2kK}}^2.
\end{array}\nonumber
\end{align}
The upper limit of the sums are $1$ for $3K < d \leq 4K$ and $2$ for $4K < d \leq 5K$. For brevity, the exact expressions of $y_1$, $y_2$, and $y_3$ will not be listed here, but can be found by replacing the matrix elements with the explicit expression in Eq.~\eqref{eq:sgn-Jx-mel}. Finally, by finding the eigenvalues of the two-dimensional block, we have
\begin{equation}
\begin{aligned}
    &\mathbf{P}_K^{d}
    = \frac{1}{2}\Bigg[ 1 + \sqrt{
        \frac{z_1+z_3}{2}
        + \sqrt{ \pmqty{\frac{z_1-z_3}{2}}^2+z_2 }
    } \Bigg], \\[1.5ex]
    &\begin{array}{rl}
    \text{where}\;\; z_1 &= \sum_{k=0}^{1,2} \abs{\mel{-j+K}{\sgn(J_x)}{-j+2kK}}^2, \\[.75ex]
    z_2 &= \sum_{k=0}^{1,2} \mel{-j+K}{\sgn(J_x)}{-j+2kK}\\
        &\qquad\qquad{}\times{}\mel{-j+3K}{\sgn(J_x)}{-j+2kK}, \\[.75ex]
    z_3 &= \sum_{k=0}^{1,2} \abs{\mel{-j+3K}{\sgn(J_x)}{-j+2kK}}^2.
    \end{array}
\end{aligned}
\end{equation}

\subsection{For $5K < d \leq 7K$}
For $5K < d \leq 7K$, similar to the preceding section, we simplify each six- and seven-dimensional block into a three-dimensional one. The $-\bar{m}$ subspace of $\{\EE_K[\sgn(J_x)]\}^2$ is
\begin{equation}
\Pi_K^{(-\bar{m})}\sgn(J_x)\Pi_K^{(+\bar{m})}\sgn(J_x)\Pi_K^{(-\bar{m})}
{}\widehat{=}{}\pmqty{
y_1 & y_2 & y_4 \\
y_2 & y_3 & y_5 \\
y_4 & y_5 & y_6
},
\end{equation}
\begin{align}
\text{where}\;& y_1 = \sum_{k=0}^{3,4} \abs{\mel{\bar{m}+K}{\sgn(J_x)}{\bar{m}+2kK}}^2, \nonumber\\[.75ex]
&y_2 = \textstyle\sum_{k=0}^{3,4} \mel{\bar{m}+K}{\sgn(J_x)}{\bar{m}+2kK} \nonumber\\
    &\qquad\qquad{}\times{}\mel{\bar{m}+3K}{\sgn(J_x)}{\bar{m}+2kK}, \nonumber\\[.75ex]
&y_3 = \textstyle\sum_{k=0}^{3,4} \abs{\mel{\bar{m}+3K}{\sgn(J_x)}{\bar{m}+2kK}}^2, \nonumber\\[.75ex]
&y_4 = \textstyle\sum_{k=0}^{3,4} \mel{\bar{m}+K}{\sgn(J_x)}{\bar{m}+2kK}\nonumber\\
    &\qquad\qquad{}\times{}\mel{\bar{m}+5K}{\sgn(J_x)}{\bar{m}+2kK}, \nonumber\\[.75ex]
&y_5 = \textstyle\sum_{k=0}^{3,4} \mel{\bar{m}+3K}{\sgn(J_x)}{\bar{m}+2kK}\nonumber\\
    &\qquad\qquad{}\times{}\mel{\bar{m}+5K}{\sgn(J_x)}{\bar{m}+2kK}, \nonumber\\[.75ex]
&y_6 = \textstyle\sum_{k=0}^{3,4} \abs{\mel{\bar{m}+5K}{\sgn(J_x)}{\bar{m}+2kK}}^2.\nonumber
\end{align}
The upper limit of the sums are $3$ for $5K < d \leq 6K$ and $4$ for $6K < d \leq 7K$. Using standard tools to find the eigenvalues of a symmetric three-dimensional matrix \cite{closed-form-3d-matrix},
\begin{align*}
    & \mathbf{P}_K^{d} = \frac{1}{2}\pqty{1 + \frac{1}{\sqrt{3}}\sqrt{z_1+z_3+z_6 + 2\sqrt{u}\cos(\phi)} }, \\
    \text{where}\;&\phi = \frac{1}{3}\arctan(\sqrt{\frac{4u^3-v^2}{v}}) - \begin{cases}
        0 & \text{for $v \geq 0$} \\
        \frac{\pi}{3} & \text{for $v < 0$},
    \end{cases}\\
    &u = z_1^2 + z_3^2 + z_6^2 - z_1z_3 - z_1z_6 - z_3z_6 \\
    &\qquad{}+{} 3(z_2^2 + z_4^2 + z_5^2), \\
    &v = (2z_1-z_3-z_6)(2z_3-z_1-z_6)(2z_6-z_1-z_3) \\
        &\qquad{}+{}54z_2z_4z_5 - 9z_2^2(2z_6-z_1-z_3)\\
        &\qquad{}-{}
        9z_4^2(2z_3-z_1-z_6) -
        9z_5^2(2z_1-z_3-z_6),\\
    &z_1 = \textstyle\sum_{k=0}^{3,4} \abs{\mel{-j+K}{\sgn(J_x)}{-j+2kK}}^2, \\[.75ex]
    &z_2 = \textstyle\sum_{k=0}^{3,4} \mel{-j+K}{\sgn(J_x)}{-j+2kK} \\
        &\qquad\qquad{}\times{}\mel{-j+3K}{\sgn(J_x)}{-j+2kK}, \\[.75ex]
    &z_3 = \textstyle\sum_{k=0}^{3,4} \abs{\mel{-j+3K}{\sgn(J_x)}{-j+2kK}}^2, \\[.75ex]
    &z_4 = \textstyle\sum_{k=0}^{3,4} \mel{-j+K}{\sgn(J_x)}{-j+2kK}\\
        &\qquad\qquad{}\times{}\mel{-j+5K}{\sgn(J_x)}{-j+2kK}, \\[.75ex]
    &z_5 = \textstyle\sum_{k=0}^{3,4} \mel{-j+3K}{\sgn(J_x)}{-j+2kK}\\
        &\qquad\qquad{}\times{}\mel{-j+5K}{\sgn(J_x)}{-j+2kK}, \\[.75ex]
    &z_6 = \textstyle\sum_{k=0}^{3,4} \abs{\mel{-j+5K}{\sgn(J_x)}{-j+2kK}}^2.
\end{align*}

\subsection{\label{apd:spin-inseparable}Inseparability of the maximally-violating state across the composition of two spins}
Consider a composite system of two particles with angular momenta $j_1$ and $j_2$. Without any loss of generality, let us take $j_1 \leq j_2$. Then
\begin{equation}
J_z = J_z^{(j_1)}\otimes\mathbbm{1}_{j_2} + \mathbbm{1}_{j_1} \otimes J_z^{(j_2)}
= \bigoplus_{j=j_2-j_1}^{j_2+j_1} J_z^{(j)}.
\end{equation}
When performing the protocol with $K$ angles, if the maximum violation $\mathbf{P}_K^{K+1}$ is obtained, the state must belong in the block $j=\frac{K}{2}$ and must be of the form
\begin{equation}\label{eq:max-violating-composite}
\begin{aligned}
\ket{\Psi_K} &= \frac{1}{\sqrt{2}}\bqty{\ket{j}_j + (-1)^{\pqty{K-1}/{2}}\ket{-j}_j} \\
&= \frac{1}{\sqrt{2}}\bqty{\mathbbm{1} + (-1)^{\pqty{K-1}/{2}}e^{-i\pi J_y}}\ket{j}_j.
\end{aligned}
\end{equation}
Defining the Clebsch-Gordan coefficients as $C_{m,n}^{m+n} = \pqty{{}_{j_1}\!\bra{m}\otimes {}_{j_2}\!\bra{n}}\ket{m+n}_{j}$, the state $\ket{j}_j$ can be expanded in the $j_1$ and $j_2$ basis as
\begin{equation}
\ket{j}_j = \sum_{m=j-j_2}^{j_1} C_{m,j-m}^j \ket{m}_{j_1}\otimes\ket{j-m}_{j_2}.
\end{equation}
The upper and lower limits of the sum come from the requirement that $m \leq j_1$ and $j-m \leq j_2$. With this, Eq.~\eqref{eq:max-violating-composite} is
\begin{widetext}
\begin{equation}\label{eq:max-violating-ind}
\begin{aligned}
\ket{\Psi_K} &=
\frac{1}{\sqrt{2}}\pqty{
    \mathbbm{1} + (-1)^{\pqty{K-1}/{2}} e^{-i\pi J_y^{(1)}} e^{-i\pi J_y^{(2)}}
} \sum_{m=j-j_2}^{j_1} C_{m,j-m}^j \ket{m}_{j_1}\otimes\ket{j-m}_{j_2} \\
&= \frac{1}{\sqrt{2}}\pqty\Big{\sum_{m=j-j_2}^{j_1} C_{m,j-m}^j \ket{m}_{j_1}\otimes\ket{j-m}_{j_2} + (-1)^{j_1+j_2-\frac{1}{2}}\sum_{m=-j_1}^{j_2-j} C_{-m,j+m}^j \ket{m}_{j_1}\otimes\ket{-m-j}_{j_2}}.
\end{aligned}
\end{equation}
\end{widetext}
\subsubsection{Proof of Inseparability}
Let us assume that the maximally violating state is separable, that is, $\ket{\Psi_K}=\ket{\psi_1}_{j_1}\otimes\ket{\psi_2}_{j_2}$. Then
\begin{widetext}
\begin{equation}
\begin{aligned}
\pqty{\bra{j_1}_{j_1}\otimes\mathbbm{1}_{j_2}}\ket{\Psi_K} &= \braket{j_1}{\psi_1}_{j_1} \; \ket{\psi_2}_{j_2} \\
&= \frac{1}{\sqrt{2}} C_{j_1,j-j_1}^j \ket{j-j_1}_{j_2} + (-1)^{j_1+j_2-\frac{1}{2}} \frac{1}{\sqrt{2}}C_{-j_1,j+j_1}^j \ket{-j_1-j}_{j_2}, \\
\pqty{\bra{j-j_2}_{j_1}\otimes\mathbbm{1}_{j_2}}\ket{\Psi_K} &= \braket{j-j_2}{\psi_1}_{j_1} \; \ket{\psi_2}_{j_2} \\
&= 
\frac{1}{\sqrt{2}} C_{j-j_2,j_2}^j \ket{j_2}_{j_2} + (-1)^{j_1+j_2-\frac{1}{2}} \frac{1}{\sqrt{2}}C_{j_2-j,2j-j_2}^j \ket{j_2-2j}_{j_2}.
\end{aligned}
\end{equation}
\end{widetext}
Note that $C_{j_1,j-j_1}^j$ and $C_{j-j_2,j_2}^j$ are always nonzero for $j_2-j_1 \leq j \leq j_2+j_1$, while $C_{-j_1,j+j_1}^j$ is nonzero only when $j=j_2-j_1$ and $C_{j_2-j,2j-j_2}^j$ is nonzero only when $j_2-j_1 \leq j \leq j_2$. Hence, there are always terms on the right-hand side that are nonzero, which implies that both $\braket{j_1}{\psi_1}_{j_1}$ and $\braket{j-j_2}{\psi_1}_{j_1}$ must be nonzero. Therefore, we can divide the equation by these terms to obtain an expression for $\ket{\psi_2}_{j_2}$ and compare these expressions for the different cases of $j_1$ and $j_2$.

For the case $j_2 < j \leq j_1 + j_2$,
\begin{equation*}
\begin{aligned}
\ket{\psi_2}_{j_2}
&= \frac{1}{\sqrt{2}\braket{j_1}{\psi_1}_{j_1}} C_{j_1,j-j_1}^j \ket{j-j_1}_{j_2}, \\
\ket{\psi_2}_{j_2}
&= \frac{1}{\sqrt{2}\braket{j-j_2}{\psi_1}_{j_1}} C_{j-j_2,j_2}^j \ket{j_2}_{j_2}.
\end{aligned}
\end{equation*}
These two expressions for $\ket{\psi_2}_{j_2}$ can be consistent only when $\ket{j-j_1}_{j_2}=\ket{j_2}_{j_2}$. This cannot be satisfied when $j_2 < j < j_1 + j_2$, and in those cases $\ket{\Psi_K} \neq \ket{\psi_1}_{j_1}\otimes\ket{\psi_2}_{j_2}$.

Meanwhile, for the case $j=j_1+j_2$, we can write out $\ket{\Psi_K}$ explicitly:
\begin{equation*}
    \ket{\Psi_K} = \frac{1}{\sqrt{2}}\bqty{
        \ket{j_1}_{j_1}\otimes\ket{j_2}_{j_2} + (-1)^{\pqty{K-1}/{2}}\ket{-j_1}_{j_1}\otimes\ket{-j_2}_{j_2}
    }.
\end{equation*}
This state is clearly entangled. Therefore, $\ket{\Psi_K}$ is entangled when $j_2 < j \leq j_1 + j_2$.

For the case $j_2-j_1 < j \leq j_2$,
\begin{equation*}
\begin{aligned}
\ket{\psi_2}_{j_2}
&= \frac{1}{\sqrt{2}\braket{j_1}{\psi_1}_{j_1} } C_{j_1,j-j_1}^j \ket{j-j_1}_{j_2}, \\
\ket{\psi_2}_{j_2}
&= \frac{1}{\sqrt{2}\braket{j-j_2}{\psi_1}_{j_1}}\Big[ C_{j-j_2,j_2}^j \ket{j_2}_{j_2} \\
&\qquad\qquad{}+{} (-1)^{j_1+j_2-\frac{1}{2}} C_{j_2-j,2j-j_2}^j \ket{j_2-2j}_{j_2}\Big].
\end{aligned}
\end{equation*}
The two expressions for $\ket{\psi_2}_{j_2}$ are clearly contradictory, so $\ket{\Psi_K}\neq \ket{\psi_1}_{j_1} \otimes \ket{\psi_2}_{j_2}$ for $j_2-j_1 < j \leq j_2$.

For the case $j = j_2-j_1$,
\begin{equation*}
\begin{aligned}
\ket{\psi_2}_{j_2}
&= \frac{1}{\sqrt{2}\braket{j_1}{\psi_1}_{j_1}} \Big[
C_{j_1,j-j_1}^j \ket{j-j_1}_{j_2} \\
&\qquad\qquad{}+{} (-1)^{j_1+j_2-\frac{1}{2}} C_{-j_1,j+j_1}^j\ket{-j_1-j}_{j_2}\Big], \\
\ket{\psi_2}_{j_2}
&= \frac{1}{\sqrt{2}\braket{j-j_2}{\psi_1}_{j_1}} \Big[
C_{j-j_2,j_2}^j \ket{j_2}_{j_2} \\
&\qquad\qquad{}+{} (-1)^{j_1+j_2-\frac{1}{2}} C_{j_2-j,2j-j_2}^j \ket{j_2-2j}_{j_2}\Big].
\end{aligned}
\end{equation*}
In this case, the expressions can be equal only if $\ket{j-j_1}_{j_2}=\ket{j_2}_{j_2}$ and $\ket{-j_1}_{j_2}=\ket{j_2-2j}_{j_2}$ or if $\ket{j-j_1}_{j_2}=\ket{j_2-2j}_{j_2}$ and $\ket{-j_1-j}_{j_2}=\ket{j_2}_{j_2}$. Every one of these conditions contradicts $j=j_2-j_1$. Therefore, $\ket{\Psi_K}$ is entangled in all four cases.

As a particular case, if the smallest spin is $j_1=\frac{1}{2}$, then $\ket{\Psi_K}$ is maximally entangled. Indeed, from \eqref{eq:max-violating-ind} we have
\begin{align*}
    \ket{\Psi_K} &=
    \frac{1}{\sqrt{2}}\ket{-\tfrac{1}{2}}_{j_1}\otimes\Big[
        C_{-\frac{1}{2},j+\frac{1}{2}}^j \ket{j+\tfrac{1}{2}}_{j_2} \\
        &\qquad\qquad{}+{}
        (-1)^{j_2} C_{\frac{1}{2},j-\frac{1}{2}}^j \ket{-j+\tfrac{1}{2}}_{j_2}
    \Big] \\
    &\quad{}+{} \frac{1}{\sqrt{2}}\ket{\tfrac{1}{2}}_{j_1}\otimes\Big[
        C_{\frac{1}{2},j-\frac{1}{2}}^j \ket{j-\tfrac{1}{2}}_{j_2} \\
        &\qquad\qquad{}+{}
        (-1)^{j_2} C_{-\frac{1}{2},j+\frac{1}{2}}^j \ket{-j-\tfrac{1}{2}}_{j_2}
    \Big],
\end{align*}
where indeed $\abs{C_{-\frac{1}{2},j+\frac{1}{2}}^j}^2 +\abs{C_{\frac{1}{2},j-\frac{1}{2}}^j}^2=1$ for all $j$. Therefore, $\rho_{j_1}=\frac{1}{2}\mathbbm{1}_{j_1}$.

\section{Results for the harmonic oscillator}\label{apd:ho}

Most of the structure found for finite dimensional systems carries over to the quantum harmonic oscillator by replacing $\ket{m_z} \to \ket{n}$, $J_z \to a^\dag a$, and $\pos(J_x) \to \pos(X)$, where
\ba \pos(X) = \int_{0}^\infty\dd{x} \ketbra{x}&=& \frac{1}{2}\bqty{\mathbbm{1}+\sgn(X)}\,.\ea In this appendix we first list those properties without repeating the proofs. Then we establish the equivalence of $\sgn{J_x}$ and $\sgn(X)$ as $j\rightarrow\infty$ and derive a lower bound for $\mathbf{P}_K^\infty$.

\subsection{Block-diagonal structure enforced by $\EE_K$}

For an arbitrary operator $A^\infty$ in Hilbert space,
\begin{equation}
    \EE_K[A^\infty] = \bigoplus_{\bar{n}=0}^{K-1} \Pi_K^{(\bar{n})}A^\infty\Pi_K^{(\bar{n})},
\end{equation}
where $\Pi_K^{(\bar{n})} \equiv \sum_{k=0}^{\infty} \ketbra{\bar{n}+kK}$. We further split $\Pi_K^{(\bar{n})}$ into even- and odd-subspace projectors: $\Pi_K^{(\bar{n})} = \Pi_K^{(+\bar{n})} + \Pi_K^{(-\bar{n})}$, where
\begin{equation}
\begin{aligned}
    \Pi_K^{(+\bar{n})} &\equiv \sum_{k=0}^{\infty} \ketbra{\bar{n}+2kK},\\
    \Pi_K^{(-\bar{n})} &\equiv \sum_{k=0}^{\infty} \ketbra{\bar{n}+(2k+1)K}.
\end{aligned}
\end{equation}

\subsection{Properties and matrix elements of $\sgn(X)$}

The operator $\sgn(X)$ transforms under the parity transformation $e^{i\pi a^\dag a}$ as
\begin{equation}\label{eq:parity-X}
    e^{i\pi a^\dag a}\sgn(X)e^{-i\pi a^\dag a} = -\sgn(X)\,.
\end{equation}
This entails the following properties: (i) $\mel{n}{\sgn(X)}{n'}=0$ when $n$ and $n'$ share the same parity and (ii) if $\lambda$ is an eigenvalue of $\EE_K[\sgn(X)]$, so is $-\lambda$. Based on these, we can prove that $\mathbf{P}_K^\infty = \frac{1}{2}(1+\sqrt{\lambda_2})$, where $\lambda_2$ is the maximum eigenvalue of $(\EE_K[\sgn(X)])^2$ over just the even blocks (or equivalently odd, but here the analytical form happens to be nicer in the even case):
\begin{equation}
    \lambda_2 = 
    \hspace{-0.75em}\max_{\begin{array}{c}\bar{n},\ket{\psi},\\\scriptstyle\braket{\psi}=1\end{array}}\hspace{-0.75em}
    \bra{\psi}\Pi_K^{(+\bar{n})}\sgn(X)\Pi_K^{(-\bar{n})}\sgn(X)\Pi_K^{(+\bar{n})}\ket{\psi}\,.
\end{equation} As the matrix elements of $\bra{n}\sgn(X)\ket{n'}$ are zero when $n'$ and $n$ are both even or both odd, throughout the rest of the section we will consider only $n'$ odd and $n$ even. By first expanding $\sgn(X)$ in the position basis,
\begin{equation}\label{eq:sgn-x-setup}
\begin{aligned}
    &\mel{n}{\sgn(X)}{n'} \\
    &= \int_{0}^\infty\dd{x} \pqty{
        \braket{n}{x}\!\braket{x}{n'}
        - \braket{n}{-x}\!\braket{-x}{n'}
    } \\
    &= 2\int_{0}^\infty\dd{x}
        \braket{n}{x}\!\braket{x}{n'},
\end{aligned}
\end{equation}
where we have used $\braket{-x}{n} = \braket{x}{n}$ for $n$ even and  $\braket{-x}{n'} = -\braket{x}{n'}$ for $n'$ odd. These wave functions are well known and are given by
\begin{align}
    \braket{x}{n} &= \frac{2^{-\frac{n}{2}}e^{-\frac{x^2}{2}}}{\sqrt{n!\sqrt{\pi}}}H_{n}(x), \label{eq:wavefunction-even}\\
    \braket{x}{n'} &= \frac{2^{-\frac{n'}{2}}e^{-\frac{x^2}{2}}}{\sqrt{n'!\sqrt{\pi}}}H_{n'}(x), \label{eq:wavefunction-odd}
\end{align}
where $H_{n}(x)$ and $H_{n'}(x)$ are the Hermite polynomials
\begin{align}
H_{n}(x) &= n! (-1)^{-\frac{n}{2}}
\sum_{k=0}^{\frac{n}{2}} \frac{(-1)^k \pqty{2x}^{2k}}{(2k)!\pqty{\frac{n}{2}-k}!}, \label{eq:hermite-even}\\
H_{n'}(x) &= n'! (-1)^{\frac{n'-1}{2}}
\sum_{k=0}^{\frac{n'-1}{2}} \frac{(-1)^k \pqty{2x}^{2k+1}}{(2k+1)!\pqty{\frac{n'-1}{2}-k}!}\label{eq:hermite-odd}.
\end{align}
Then $\mel{n}{\sgn(X)}{n'} = \mel{n'}{\sgn(X)}{n}$ can be evaluated by substituting Eqs.~\eqref{eq:wavefunction-even}--\eqref{eq:hermite-odd} into Eq.~\eqref{eq:sgn-x-setup},
\begin{widetext}
\begin{align}
\mel{n}{\sgn(X)}{n'}
&=2\int_{0}^\infty\dd{x}
\frac{2^{-\frac{n}{2}}e^{-\frac{x^2}{2}}}{\sqrt{n!\sqrt{\pi}}}\pqty{
        n! (-1)^{-\frac{n}{2}}
        \sum_{k=0}^{\frac{n}{2}} \frac{(-1)^k}{(2k)!\pqty{\frac{n}{2}-k}!}\pqty{2x}^{2k}
} \nonumber\\
&\qquad\qquad\times \frac{2^{-\frac{n'}{2}}e^{-\frac{x^2}{2}}}{\sqrt{n'!\sqrt{\pi}}}
\pqty{
    n'! (-1)^{\frac{n'-1}{2}}
    \sum_{k'=0}^{\frac{n'-1}{2}} \frac{(-1)^{k'}}{(2k'+1)!\pqty{\frac{n'-1}{2}-k'}!}\pqty{2x}^{2k'+1}
}  \nonumber\\
&= (-1)^{\frac{n'-n-1}{2}} 2^{-\pqty{\frac{n'+n}{2}-1}}
    \sqrt{\frac{n!n'!}{\pi}}
    \sum_{k=0}^{\frac{n}{2}}
    \sum_{k=0}^{\frac{n'-1}{2}}
    \frac{(-1)^{k+k'}2^{2(k+k')}}{(2k)!(2k'+1)!\pqty{\frac{n}{2}-k}!\pqty{\frac{n'-1}{2}-k'}!}
    \underbrace{\int_0^{\infty}\dd{x^2} e^{-x^2} (x^2)^{k+k'}}_{(k+k')!} \nonumber\\
&= \frac{(-1)^{\frac{n'-n-1}{2}} 2^{-\pqty{\frac{n'+n}{2}-1}}}{n'-n}
\sqrt{
    \frac{n'}{\pi}
    \pmqty{
        n\\
        \frac{n}{2}
    }
    \pmqty{
        n'-1\\
        \frac{n'-1}{2}
    }
}\qquad\text{for $n$ even, $n'$ odd.}\label{eq:sgn-x-mel}
\end{align}
\end{widetext}

\subsection{Equivalence of $\sgn(J_x)$ and $\sgn(X)$ as $j\to\infty$}
While the operator $J_x(t) = J_x\cos(\omega t) + J_y\sin(\omega t)$ that we considered in the main text evolves similarly to the position operator $X(t) = X\cos(\omega t) + P\sin(\omega t)$ considered by Tsirelson, these operators are not exactly equivalent as $\comm{X}{P} = i\mathbbm{1}$ while $\comm{J_x}{J_y} = iJ_z$. Nonetheless, we will show in this section that $\sgn(J_x) \to \sgn(X)$ as $j \to \infty$, which means that $\mathbf{P}_K^d \to \mathbf{P}_K^\infty$ in that limit.

We relabel the angular momentum states $\ket{\tilde{n}} \equiv \ket{m_z = \tilde{n}-j}$ such that $\ket{\tilde{n}=0}$ refers to the lowest-energy state of $H^d=\omega J_z$ and $\ket{\tilde{n}}$ refers to the $\tilde{n}$th excited state. This is analogous to $\ket{n}$ being the $\tilde{n}$th excited state of $H^\infty = \hbar\omega a^\dag a$.

Consider $\mel{\tilde{n}}{\sgn(J_x)}{\tilde{n}'}$, where $\tilde{n}'=j+m'$ is odd and $\tilde{n}=j+m$ is even. Then
\begin{widetext}
\begin{align}
&\mel{\tilde{n}}{\sgn(J_x)}{\tilde{n}'} \nonumber\\
&= \frac{(-1)^{\frac{\tilde{n}'-\tilde{n}-1}{2}}2^{-(2j-1)}}{\tilde{n}'-\tilde{n}}
\sqrt{
\tilde{n}'
\pmqty{\tilde{n}\\\frac{\tilde{n}}{2}}
\pmqty{\tilde{n}'-1\\\frac{\tilde{n}'-1}{2}}
}\times\begin{cases}
\sqrt{
    (2j-\tilde{n}')
    \pmqty{2j-\tilde{n}\\\frac{2j-\tilde{n}}{2}}
    \pmqty{2j-\tilde{n}'-1\\\frac{2j-\tilde{n}'-1}{2}}
} & \text{$d$ odd},\\
\sqrt{
    (2j-\tilde{n})
    \pmqty{2j-\tilde{n}-1\\\frac{2j-\tilde{n}-1}{2}}
    \pmqty{2j-\tilde{n}'\\\frac{2j-\tilde{n}'}{2}}
} & \text{$d$ even},
\end{cases} \nonumber\\
&= \frac{(-1)^{\frac{\tilde{n}'-\tilde{n}-1}{2}}2^{-\pqty{\frac{\tilde{n}'+\tilde{n}}{2}-1}}}{\tilde{n}'-\tilde{n}}
\sqrt{
\tilde{n}'
\pmqty{\tilde{n}\\\frac{\tilde{n}}{2}}
\pmqty{\tilde{n}'-1\\\frac{\tilde{n}'-1}{2}}
} \label{eq:sgn-Jx-mel-lim}\\
&\hspace{13.5em}{}\times{}\begin{cases}
\sqrt{
    \pqty{1-\frac{\tilde{n}'}{j}}
    \times
    \sqrt{j} 2^{-(2j-\tilde{n})}
    \pmqty{2j-\tilde{n}\\\frac{2j-\tilde{n}}{2}}
    \times
    \sqrt{j} 2^{-(2j-\tilde{n}'-1)}
    \pmqty{2j-\tilde{n}'-1\\\frac{2j-\tilde{n}'-1}{2}}
} & \text{$d$ odd},\\
\sqrt{
    \pqty{1-\frac{\tilde{n}}{j}}
    \times
    \sqrt{j} 2^{-(2j-\tilde{n}-1)}
    \pmqty{2j-\tilde{n}-1\\\frac{2j-\tilde{n}-1}{2}}
    \times
    \sqrt{j} 2^{-(2j-\tilde{n}')}
    \pmqty{2j-\tilde{n}'\\\frac{2j-\tilde{n}'}{2}}
} & \text{$d$ even}.
\end{cases}
\end{align}
\end{widetext}
The dependence of these matrix elements on $j$ is mostly of the form 
\begin{equation}\label{eq:stirling-part-1}
    \sqrt{j}2^{-(2j-y)}\pmqty{2j-y\\\frac{2j-y}{2}} = \frac{1}{\sqrt{\pi(1-\frac{y}{2j})}} e^{\gamma}
\end{equation}
for some $\gamma$, with $y$ replaced by $\tilde{n}$, $\tilde{n}-1$, $\tilde{n}'$ or $\tilde{n}'-1$ for the corresponding expression in Eq.~\eqref{eq:sgn-Jx-mel-lim}. The right-hand side of Eq.~\eqref{eq:stirling-part-1} is due to Stirling's formula, where $\gamma$ has an upper and lower bound \cite{stirling-bound}
\begin{equation}
\begin{gathered}
\frac{1}{12(2j-y)+1} - \frac{1}{3(2j-y)} \leq \gamma, \\
\gamma \leq \frac{1}{12(2j-y)} - \frac{2}{6(2j-y)+1}.
\end{gathered}
\end{equation}
As $j\to\infty$, both the lower and upper bounds of $\gamma$ approaches 0, so Eq.~\eqref{eq:stirling-part-1} approaches $1/\sqrt{\pi}$. Hence,
\begin{equation}
\begin{aligned}
&\lim_{j\to\infty} \mel{\tilde{n}}{\sgn(J_x)}{\tilde{n}'}\\
&\quad{}={}  \frac{(-1)^{\frac{\tilde{n}'-\tilde{n}-1}{2}}2^{-\pqty{\frac{\tilde{n}'+\tilde{n}}{2}-1}}}{\tilde{n}'-\tilde{n}}
\sqrt{
\frac{\tilde{n}'}{\pi}
\pmqty{\tilde{n}\\\frac{\tilde{n}}{2}}
\pmqty{\tilde{n}'-1\\\frac{\tilde{n}'-1}{2}}
} \\
&\quad{}={} \mel{n = \tilde{n}}{\sgn(X)}{n' = \tilde{n}'}.
\end{aligned}
\end{equation}
The converse case for $\tilde{n}$ odd and $\tilde{n}'$ even can be found from $\sgn(J_x) = \sgn(J_x)^\dag$, while all other matrix elements are zero, so $\lim_{j\to\infty}\sgn(J_x) = \sgn(X)$. Therefore, $\mathbf{P}^\infty_K$, in both its original formulation and the angular momentum system as $j\to\infty$, are the same.

\subsection{Closed-form lower bound for $\mathbf{P}^\infty_K$}
Let $\lambda_2$ be the maximum eigenvalue of $\{\EE_{K}\bqty{\sgn(X)}\}^2$. Since $\lambda_2 \geq \mel{\psi}{(\EE_{K}\bqty{\sgn(X)})^2}{\psi}$ for $\braket{\psi} = 1$, any normalized state $\ket{\psi}$ provides us with a lower bound for $\lambda_2$, hence for $\mathbf{P}^\infty_K = \frac{1}{2}(1+\sqrt{\lambda_2})$.

From the block-diagonal structure of $\{\EE_{K}\bqty{\sgn(X)}\}^2$, we can consider just the even states. In addition, from the first few finite-dimensional cases given in appendix~\ref{sec:particular-values} and from solving for the eigenvectors of $Q_K^\infty$ in a truncated Hilbert space, we deduce that the eigenvector of $\{\EE_{K}\bqty{\sgn(X)}\}^2$ will have a large component of $\ket{0}$. Hence, we study the lower bound for $\ket{\psi}=\ket{0}$:
\begin{equation}\label{eq:lower-bound-step-1}
\begin{aligned}
\lambda_2 &\geq
\sum_{k=0}^\infty \mel{0}{\sgn(X)}{(2k+1)K}\!\!\mel{(2k+1)K}{\sgn(X)}{0} \\
&= \frac{2}{\pi} \sum_{k=0}^\infty \frac{ 2^{-(2k+1)K}}{(2k+1)K}
    \pmqty{
        (2k+1)K-1\\
        \frac{(2k+1)K-1}{2}
    }.
\end{aligned}
\end{equation}
Equation \eqref{eq:lower-bound-step-1} is reminiscent of the series expansion
\begin{equation}\label{eq:lower-bound-step-2}
\frac{2}{\pi\sqrt{x}}\arcsin(\sqrt{x})
= \frac{2}{\pi} \sum_{k=0}^\infty \frac{ 2^{-(2k+1)}}{2k+1}
\pmqty{
    2k\\
    k
}x^k.
\end{equation}
In fact, we recognize the series in Eq.~\eqref{eq:lower-bound-step-1} to be a subseries of the right-hand side of Eq.~\eqref{eq:lower-bound-step-2} with $x=1$ and values of $k$ such that $k\bmod K = \frac{K-1}{2}$. Given that the full series converges to some function of $x$, the subseries can be expressed in terms of this function as \cite{closed-form-subseries}
\begin{equation}
\begin{aligned}
\lambda_2 &\geq \frac{1}{K} \sum_{k=0}^{K-1}e^{-i\frac{2\pi k}{K}\pqty{\frac{K-1}{2}}}
\frac{2}{\pi \sqrt{e^{i\frac{2\pi k}{K}}}} \arcsin(\sqrt{e^{i\frac{2\pi k}{K}}}) \\
&= \frac{1}{K} + \frac{2}{\pi K} \sum_{k=1}^{\pqty{K-1}/{2}} (-1)^k\arccos\bqty{2\sin(\frac{\pi k}{K})-1}.
\end{aligned}
\end{equation}
The simplification was performed by expressing $\arcsin$ in terms of the complex . In particular, for the case $K=3$, we have
\begin{equation}
    \lambda_2 \geq \frac{1}{3}\pqty{1 - \frac{2}{\pi} \arccos(\sqrt{3}-1)} \approx 0.174,
\end{equation}
and therefore $\mathbf{P}_K^\infty \geq 0.7087$.

\end{appendix}
\end{document}